# Econometric Analysis of Pandemic Disruption and Recovery Trajectory in the U.S. Rail Freight Industry

**Max T.M. Ng**
Graduate Research Assistant
Transportation Center
Northwestern University
600 Foster Street
Evanston, IL 60208, USA
Email: maxng@u.northwestern.edu

**Hani S. Mahmassani**
William A. Patterson Distinguished Chair in Transportation
Director, Transportation Center
Northwestern University
600 Foster Street
Evanston, IL 60208, USA
Email: masmah@northwestern.edu

**Joseph L. Schofer**
Professor Emeritus of Civil and Environmental Engineering
Northwestern University
2145 Sheridan Road, Evanston, IL 60208, USA
Email: j-schofer@northwestern.edu

**ABSTRACT**

To measure the impacts on U.S. rail and intermodal freight by economic disruptions of the 2007-09 Great Recession and the COVID-19 pandemic, this paper uses time series analysis with the AutoRegressive Integrated Moving Average (ARIMA) family of models and covariates to model intermodal and commodity-specific rail freight volumes based on pre-disruption data. A framework to construct scenarios and select parameters and variables is demonstrated. By comparing actual freight volumes during the disruptions against three counterfactual scenarios, Trend Continuation, Covariate-adapted Trend Continuation, and Full Covariate-adapted Prediction, the characteristics and differences in magnitude and timing between the two disruptions and their effects across nine freight components are examined.

Results show the disruption impacts differ from measurement by simple comparison with pre-disruption levels or year-on-year comparison depending on the structural trend and seasonal pattern. Recovery Pace Plots are introduced to support comparison in recovery speeds across freight components. Accounting for economic variables helps improve model fitness. It also enables evaluation of the change in association between freight volumes and covariates, where intermodal freight was found to respond more slowly during the pandemic, potentially due to supply constraint.

# 1. INTRODUCTION

The COVID-19 pandemic and the measures to contain it brought large and rapid changes to the U.S. freight industry in 2020. Railroads, as the backbone of the economy, saw freight volumes drop sharply by 18% from February to April 2020, but experienced substantial recovery within five months. The impacts of this supply shock, brought by public health concerns and lockdown measures, were significant and uneven across markets, and the characteristics of changes were markedly different from the last major economic disruption, the Great Recession in 2007-2009, which was a more longer-term demand shock due to reduction in income and economic activities. (Schofer et al., 2021) Rail intermodal (IM) and carload freight are critical components of the supply chain, the former moving finished goods and manufacturing inputs, the latter bulk commodities such a grain, coal, and enegy products. Freight volumes indicate not only the levels of economic activities during disruptions, but also the service constraints and resilience, from which insights can be drawn to inform resilience preparation and investment decisions.

The U.S. Department of Transportation (2022) published a report on supply chain assessment in February 2022, stressing the importance of supply chain resilience to the greater economy and society. It identified data availability and knowledge gaps as a critical challenge, while existing freight models are "*inadequate to permit detailed analysis of supply chains of effects of disruptions*" (p32). This paper aims to address these gaps.

Existing measurement methodologies of disruption impacts and the subsequent recovery are insufficient. Direct comparison with pre-disruption levels is an intuitively appealing and conceptually simple approach, but it is prone to interference of seasonal impacts. Rail freight volume is highly seasonal, influenced by industry practices and socio-economic activities such as sales in year-end holidays and reduction in operations between Christmas and New Year. IM freight traffic in particular, tends to increase at the retail peak seasons and drop during year-end holidays. Furthermore, the rail freight sector and the economy it supports possess a certain dynamic driven by both external and internal factors, with structural trends in terms of the spatio-temporal patterns of the demand for various commodities and products. For example, pre-pandemic, coal volumes had been in a secular decline, while IM traffic had been in a general uptrend - albeit with fluctuations driven by economic and trade forces. While examining year-on-year changes might address the seasonality issue, assessment of the disruption and especially the recovery remains incomplete without consideration of the prevailing pre-pandemic structural trends and their counterfactual continuation. Capturing these trends, and projecting them forward provide an additional benchmark against which to assess the recovery and help predict its trajectory.

This leads to three research questions (RQs) about how such impacts and the trajectory of the recovery may be quantified, compared, and projected:

**RQ1**: How can the disruption impacts be measured in rail freight volumes while separating freight activities from seasonality and structural trends?

**RQ2**: Does the inclusion of broad economic metrics (covariates) in model specifications improve the ability to capture temporal patterns of rail freight volumes and lead to more meaningful pre- and post- pandemic comparisons?

**RQ3**: Did the association between freight volumes and the underlying economic metrics (covariates) established pre-pandemic continue to hold during the disruption and subsequent recovery - *i.e.* do the actual values of these exogenous economic indicators still help forecast the recovery in rail freight volumes?

To address these questions, this paper presents an econometric analysis using time-series models of rail freight volumes pre-disruption and applies them to project post-disruption counterfactual baselines against which to assess the impact of the disruption, the extent of the recovery, and its likely future direction. The models are developed and applied following three distinct methodological scenarios, with respect to the above three RQs. **Figure 1** summarizes the three approaches, scenarios constructed, and corresponding RQs.

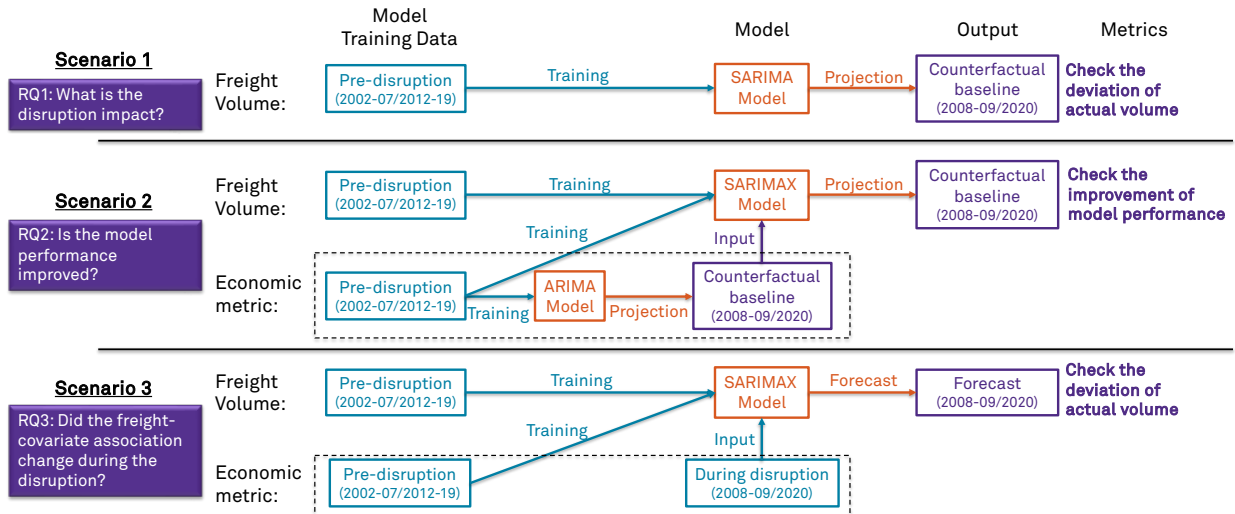

**Figure 1 – Flowchart for Scenario Construction corresponding to Research Questions (RQs)**

Scenario 1 - Trend Continuation, is constructed by Seasonal AutoRegressive Integrated Moving Average (SARIMA) models trained by the pre-disruption data series of the freight volume variables of interest. The models then project counterfactual baselines that extend the underlying pre-pandemic trend. The difference between the actual freight volume and that counterfactual baseline is then interpreted as an estimate of the impact brought by the disruption.

Scenario 2 - Covariate-adapted Trend Continuation, is the enhanced result of Seasonal AutoRegressive Integrated Moving Average with eXogenous regressors (SARIMAX) models created by selecting and incorporating economic indicators as covariates. The main models are trained with pre-disruption freight volume and economic metrics. By taking the covariates projected during a disruption in a separate model as the input, the main models then construct a new counterfactual against which to assess the pandemic impact and recovery status. The improvement in prediction quality suggests whether the incorporation of the economic metric is beneficial.

Scenario 3 - Actual Covariate-adapted Forecast, is the output of the same SARIMAX model in Scenario 2, yet with the actual data series of the economic indicators during the disruption period to forecast freight volume under disruption given the economic situation reflected by the covariate. Discrepancies between the forecasted values and the actual volume would suggest possible changes in their pattern of association, along with the influence of other constraints and factors.

Taken together, the three scenarios delineate a range of counterfactual baselines to assess the pandemic impact and its recovery trajectory. They also provide a basis for forecasting the likely direction of the recovery. For clarity, "*prediction*" refers to the output of time series model exclusively. "*Projection*" refers to extension of past trends (based on pre-disruption freight volumes in Scenario 1, and also pre-disruption economic metrics in Scenario 2) and "*forecast*" refers to prediction of future freight volumes (based on pre-disruption freight volumes and economic metrics pre- and during disruption in Scenario 3).

The approach is applied to the two recent major economic disruptions to the freight industry, the Great Recession (2008-2009) and the COVID-19 pandemic (2020). Freight volume data for IM and eight commodities, which cover around 90% of total rail freight traffic in the U.S., were evaluated under the three scenarios, focusing on the examples of IM and coal freight in 2020. To facilitate comparison across IM and commodities, Recovery Pace Plots, which correlate the magnitudes of disruption and recovery, were introduced to highlight the differences in recession and rebound/recovery patterns.

The paper continues with a review of previous efforts to analyze the impacts of economic disruptions on freight, as well as modeling freight demand particularly with time series analysis. In Section 3, it introduces the methodology, including data collection and the SARIMA/SARIMAX models,. After presenting and discussing the model results under the two economic disruptions in Section 4, it offers interpretations and conclusions in Section 5.

## 2. LITERATURE REVIEW

This section first reviews recent studies into impacts of disruptions on freight activities with a focus on rail, followed by previous research effort in time series analysis in freight or supply chain worldwide.

### 2.1 Impacts of Disruptions on Freight

Literature on impacts of short-term disruptions on rail freight is relatively limited, especially studies for the recent and evolving COVID-19 pandemic, which are likely still under development. Tardivo et al. (2021) provided a review and forecast of overall COVID-19 impact on passenger and freight rail as of 2020. Borca et al. (2021) conducted a literature review regarding recent relevant crises in Europe including COVID-19, pointing out the existing research gap on the effects of disruptions on transport modes in general.

More papers looked at initial impacts of COVID-19 pandemic beyond rail. Notteboom et al. (2021) suggested no correlation of volume fluctuation between the Great Recession and the initial state of the COVID-19 pandemic with analysis of shipping volume. Cui et al. (2021) analyzed the overall freight volume in China during the pandemic with macroscopic analysis and a general equilibrium model, concluding that the rail sectors were mainly impacted by supply-side shocks. Gudmundsson et al. (2021) made use of ARIMA with exogenous variables to estimate recovery time in air freight. Schofer et al. (2022) qualitatively analyzed the U.S. supply chain with interviews and business studies, identifying chokepoints in the IM system as containers, chassis, labor, and ports.

To conclude, a research gap was identified in the systematic analysis of the impacts of disruptions on freight activities.

### 2.2 Modeling and Time Series Analysis in Freight

Modeling rail freight demand has long been an area of focus to characterize its relationship with economic indicators and provide forecasts for policy insights and decision making. Early efforts involved the use of regression with explanatory variables. Morton (1969) incorporated macroeconomic indicators such as gross national products, personal income, and rail and truck price indices to model rail carloads in separate commodities in the U.S.; Logarithmic transformation was adopted to ensure homoscedasticity. However, the level of granularity was limited to annual data, possibly due to computational constraints. Fite, et al. (2002) proposed using stepwise multiple linear regression to make predictions based on a wide range of economic indicators, screened by correlation and identification of lead time.

Time series analysis plays an important role in modeling and forecasting freight demand, with AutoRegressive Integrated Moving Average (ARIMA) models providing a flexible and robust statistical methodology (Cambridge Systematics, 1997). More recently, ARIMA was adopted to model freight volumes in different countries ((Hunt, 2003) in Estonia; (Guo et al., 2010) in China), concluding that the ARIMA model was preferable for freight short-term prediction separated from spontaneous fluctuations. Al Hajj Hassan et al. (2020) introduced the use of a Reinforcement Learning approach parallel to ARIMA over a rolling horizon for enhancing the accuracy of demand forecasts.

Apart from regression and time series analysis techniques, other proposed models include supply chain theory (Winston, 1983), spatial econometric methods (Garrido and Mahmassani, 2000), commodity flow models (Black, 1999; Cascetta, et al., 2013; Holguín-Veras and Patil, 2008; Pompigna and Mauro, 2020), delivery simulation and optimization (Comi, 2013; Zhang et al., 2008), and behavioral approaches (Ben-Akiva, et al., 2013). However, the majority of demand model methods require large datasets on the freight system and prevailing economic conditions, and therefore excessive computation time (National Academies of Sciences, Engineering, and Medicine, 2010). ARIMA models produce a good short-term forecast while demanding fewer computational resources, but they require sufficient data in a given period and an assumption that the underlying economic condition and freight trend would continue. Therefore, they are not sensitive to variations in demands and economic disruptions.

To summarize, while ARIMA has been adopted to analyze and forecast freight volume, limited focus was placed on a structural approach to explore and quantify the impact on rail and IM freight due to short-term disruptions. The capability of ARIMA to project past trends and its extension to incorporate seasonality make it a good fit in this paper to measure disruption impacts.

## 3. METHODOLOGY

The goal of this work is to compare the actual freight data against a meaningful baseline consisting of a counterfactual projection that excludes the effect of disruptions. The main tool consists of time series econometric modeling, applied under alternative scenario assumptions that define a range of possible projected futures. The main time series model forms adopted in this work consist of the ARIMA family of models; it is suitable for this application because only the near-term data are concerned, and its assumptions of continuing trends and insensitivity towards economic disruptions are appropriate for providing the counterfactual baseline of interest.

This section will first introduce the principle of ARIMA models and its extension to seasonal models and exogenous variables. Then, the procedure of model selection and parameter evaluation will be discussion, followed by the data and application scenarios. **Table 1** shows the notations of model variables and parameters.

**Table 1 - Notation of Model Variables and Parameters**

| Parameters | $p$ | Non-seasonal autoregressive (AR) order |
|---|---|---|
| | $d$ | Non-seasonal differencing order |
| | $q$ | Non-seasonal moving average (MA) order |
| | $P$ | Seasonal autoregressive (AR) order |
| | $D$ | Seasonal differencing order |
| | $Q$ | Seasonal moving average (MA) order |
| | $S$ | Number of time steps per a seasonal cycle |
| **Model** | $x_t$ | State variables at time period $t$ |
| | $y_t$ | State variables after differencing (if applicable) at time period $t$ |
| | $z_t$ | Observed variables (*i.e.* freight volume dataset) at time period $t$ |
| | $c$ | Constant term |
| | $a_i$ | Coefficient of the $i$-th autoregressive term |
| | $m_i$ | Coefficient of the $i$-th moving average term |
| | $\varepsilon_t$ | Error term of the state variables at time period $t$ |
| | $\sigma^2$ | Normal distribution variance of error terms $\varepsilon_t$ |
| | $L^k$ | Lag operator to the power $k$ |
| | $\nabla_k$ | Difference operator in periods $k$ |
| | $\phi_p(L)$ | Autoregressive (AR) operator with order $p$ |
| | $\theta_q(L)$ | Moving average (MA) operator with order $q$ |
| | $\tilde{\phi}_P(L)$ | Seasonal autoregressive (AR) operator with order $P$ |
| | $\tilde{\theta}_Q(L)$ | Seasonal moving average (MA) operator with order $Q$ |
| | $n_t$ | Covariate for SARIMAX models |
| **Results** | $ACF_t(n)$ | Autocorrelation function (ACF) with lag $n$ at time period $t$ |
| | $PACF_t(n)$ | Partial autocorrelation function (PACF) with lag $n$ at time period $t$ |
| | $P > \lvert z \rvert\ for\ exog$ | p-value for the exogenous variable regression |
| | $P(Q)$ | p-value for the Ljung–Box test |
| | $AIC$ | Akaike's Information Criterion (AIC) value |
| | $MAPE$ | Mean Absolute Percentage Error (MAPE) |
| | $MAD$ | Mean Absolute Deviation (MAD) |
| | $\hat{y}_i$ | Forecasted value of data series at time period $t$ |

## 3.1 Autoregressive Integrated Moving Average (ARIMA) Method

As discussed in Section 2, ARIMA is chosen as the method to project and forecast freight volume to suit the assumption specific to this paper - *continuing trend*. The models are constructed with reference to (Box and Jenkins, 1970; Durbin and Koopman, 2012). As the problem introduced here is univariate (freight volume), the simpler univariate equations are first introduced, followed by formal time series notations. State space models are adopted to incorporate the effects of covariates for Scenarios 2 and 3. The state variables are the underlying structural parts of freight volumes (observation variables), in the sophisticated underlying process between supply and demand, which is still subject to structural development, periodic change, and random fluctuations.

ARIMA is a combination method of autoregressive (AR), integrated (I), and moving average (MA).

The AutoRegressive Moving Average (ARMA) approach fits the time series based on the relationship of a data point with its previous terms. Eq. (1) represents *ARMA(p,q)*, where the observed variable ($z_t$) is set to the state variable ($y_t$) and $\varepsilon_t$ is the error term normally distributed with zero mean. The first term is the constant term, then the subsequent group of terms with the coefficients of $a_i$ are AR components of order $p$, and the last group of terms with the coefficients of $m_i$ are MA components of order $q$.

Rail freight traffic data is in general not a stationary time series due to the presence of long-term structural trends. This hypothesis is tested with the augmented Dicky-Fuller (ADF) test discussed in Section 3.2. ARIMA builds on ARMA by differencing (I) non-stationary time series to stationary. Eq. (2) shows the first order of differencing, where $x_t$ is the non-stationary original time series.

With the formal time series notation, Eq. (3)-(7) present the ARIMA model with lag operators (Eq. (3)) and difference operators (Eq. (4)). The AR component (Eq. (5)) and MA component (Eq. (6)) lead to the formulation of the *ARIMA(p,1,q)* process as in Eq. (7).

$$z_t = y_t = c + a_1 y_{t-1} + a_2 y_{t-2} + \cdots a_p y_{t-p} + \varepsilon_t + m_1 \varepsilon_{t-1} + m_2 \varepsilon_{t-2} + \cdots$$
$$+ m_q \varepsilon_{t-q}, \varepsilon_t \sim N(0, \sigma^2) \tag{1}$$

$$y_t = \nabla x_t = x_t - x_{t-1} \tag{2}$$

$$L^k y_t \equiv y_{t-k} \tag{3}$$

$$\nabla_k y_t \equiv y_t - y_{t-k} \equiv (1 - L^k) y_t \tag{4}$$

$$\phi_p(L) = 1 - a_1 L - a_2 L^2 - \cdots - a_p L^p \tag{5}$$

$$\theta_q(L) = 1 + m_1 L + m_2 L^2 + \cdots + m_q L^q \tag{6}$$

$$\phi_p(L) \nabla y_t = c + \theta_q(L) \varepsilon_t \tag{7}$$

The variations of rail freight are seasonal, most often yearly. To capture this seasonality, SARIMA models address the three components of a time series - the general trend is isolated by ARIMA; the cyclic fluctuations are addressed by the seasonal parts; and the random noise residual is captured by the error term. Eq. (8) shows the form of *SARIMA (p,0,q,P,0,Q,S)* where on top of ARIMA formulations, the seasonal AR and MA components are added with the period $S$. Eq. (9) applies the first order of seasonal differencing with period $S$. With the time series notations of seasonal AR and MA components in Eq. (10) and (11) respectively, the equation of *SARIMA (p,1,q,P,1,Q,S)* can be simplified to Eq. (12).

$$z_t = y_t = c + a_1 y_{t-1} + \cdots + a_p y_{t-p} + a_S y_{t-S} + a_1 a_S y_{t-S-1} + \cdots + a_{2S} y_{t-2S} + \cdots$$
$$+ a_{PS} y_{t-PS} + a_1 a_{PS} y_{t-PS-1} + \cdots + a_p a_{PS} y_{t-PS-p} + \varepsilon_t + m_1 \varepsilon_{t-1}$$
$$+ m_2 \varepsilon_{t-2} + \cdots + m_q \varepsilon_{t-q} + m_S \varepsilon_{t-S} + m_1 m_S \varepsilon_{t-S-1} + \cdots + m_{2S} \varepsilon_{t-2S} + \cdots$$
$$+ m_{QS} \varepsilon_{t-QS} + m_1 m_{QS} \varepsilon_{t-QS-1} + \cdots + m_q m_{QS} \varepsilon_{t-QS-q} \tag{8}$$

$$y_t = \nabla_s x_t = x_t - x_{t-S} \tag{9}$$

$$\tilde{\phi}_P(L) = 1 - a_S L^S - a_{2S} L^{2S} - \cdots - a_{pS} L^{PS} \tag{10}$$

$$\tilde{\theta}_Q(L) = 1 + m_S L^S + m_{2S} L^{2S} + \cdots + m_{qS} L^{QS} \tag{11}$$

$$\tilde{\phi}_P(L)\, \phi_p(L) \nabla \nabla_S y_t = c + \tilde{\theta}_Q(L)\theta_q(L)\varepsilon_t \tag{12}$$

To reflect the impact of associated economic changes, a SARIMA model can be augmented with *a priori* covariates (or exogenous variables), such as economic metrics, to form a SARIMAX model in the basis of state space method. The errors of the observations depend on economic metrics, which are inputted as covariates ($n_t$) for regression, together with the state variables of SARIMA (Eq. (13)).

$$z_t = \beta_t n_t + y_t \tag{13}$$

The time series model was implemented with Python 3.8.8 and the tsa.statespace.SARIMAX class in the statsmodels Python module (Seabold and Perktold, 2010), which estimate the model coefficients by maximum likelihood estimation method.

## 3.2 Model Selection and Fitness Evaluation

Multiple model forms with different parameters (*p,d,q,P,D,Q,S*) and data pre-processing methods are considered and evaluated based on statistical and problem-specific criteria before selecting the final model for each freight component and each scenario. The process is summarized in **Figure 2**.

At first, the general characteristics of freight volumes were evaluated with tools including seasonal decomposition charts. This confirmed seasonality for choices between SARIMA and ARIMA models.

The ADF test, a unit root test method, was conducted to reject the null hypothesis of non-stationary time series and decide whether differencing is necessary.

The choice of AR and MA orders can be determined by observing the charts of autocorrelation function (ACF) and partial autocorrelation function (PACF) of the time series. ACF refers to the linear dependence of a data point with another at a lag *n* earlier (Eq. (14)). PACF refers to a similar autocorrelation except that the linear dependencies on the data points later than a lag *n* are eliminated (Eq. (15)). In general, the ACF of an AR model tails off, while that of an MA model cuts off after lag *q*. Similarly, the PACF of an AR model cuts off after lag *p*, while that of an MA model tails off. This assisted the identification of model orders.

$$ACF_t(n) = Corr(x_t, x_{t-n}) \tag{14}$$

$$PACF_t(n) = \begin{cases} ACF_t(1) \text{ for } n = 1 \\ \dfrac{Corr(x_t, x_{t-n} | x_{t-1}, \ldots, x_{t-n+1})}{\sqrt{Var(x_t | x_{t-1}, \ldots, x_{t-n+1}) Var(x_{t-n} | x_{t-1}, \ldots, x_{t-n+1})}} \text{ for } n > 1 \end{cases} \tag{15}$$

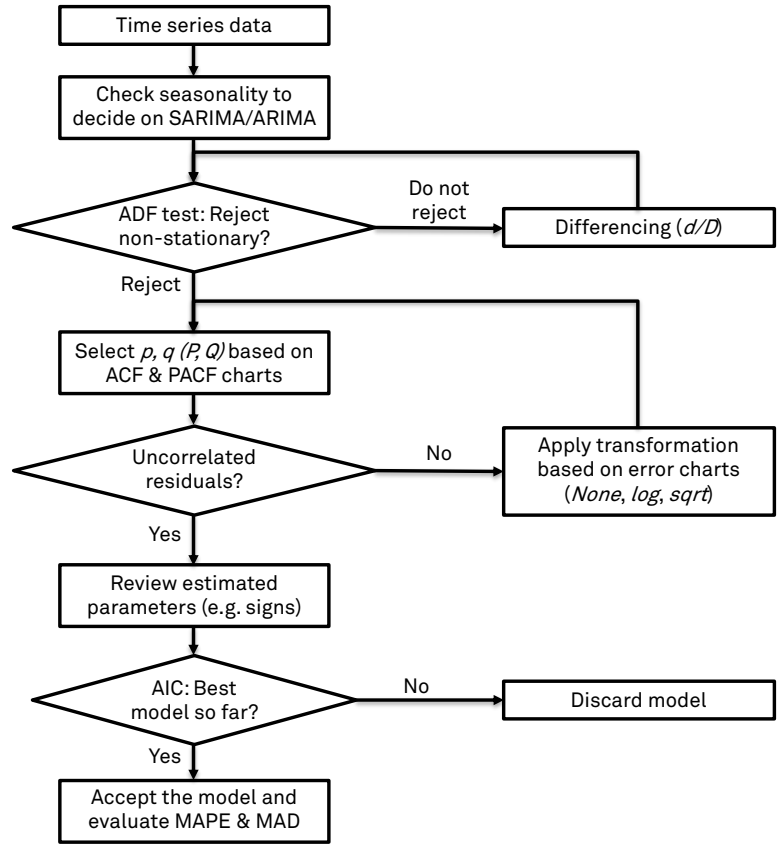

**Figure 2 – Flowchart for Model Selection and Fitness Evaluation**

The ARIMA models were then constructed with the selected parameters. An important assumption of ARIMA is the residuals are uncorrelated, *i.e.,* all meaningful information from the data series has been explained by the model variables. This is done by a multi-pronged approach. Diagnostic plots including standardized residual, error histogram, normal Q-Q plot, and correlogram assist in determining correlation among errors. A statistical method, the Ljung–Box test, which was used to reject the null hypothesis of uncorrelated residuals, served as a double check that small p-values indicate poor fitting. Uncorrelated residuals are often results of non-linear rate of change or heteroscedasticity in the data series, so transformation methods such as logarithm or square root on the original series would be carried out before applying another round of ARIMA on transformed data. Besides, standard model evaluation was carried out by examining the signs and scales of exogenous variables, AR, and MA coefficients.

For model fitness, the values of Akaike's Information Criterion (AIC) were referenced, which is defined in Eq. (16), where $k$ is the number of parameters and $LL$ is the maximum log-likelihood function of the model. A model with lower AIC is preferable, which implies it favors models with fewer parameters among others with similar likelihood.

$$AIC = 2k - 2LL \tag{16}$$

Two indicators served as the final metric to evaluate the overall fitness of the model to the past trend, Mean Absolute Percentage Error (MAPE) and Mean Absolute Deviation (MAD) (Eq.

(17) and (18)). While MAD provides the average magnitude of prediction error, MAPE acts as a standardized measure for comparison in percentages of the time series.

$$MAPE = \frac{1}{n}\sum_{i=1}^{n}\frac{|y_i - \hat{y}_i|}{y_i} \times 100\% \tag{17}$$

$$MAD = \frac{1}{n}\sum_{i=1}^{n}|y_i - \hat{y}_i| \tag{18}$$

### 3.3 Data and Application Scenarios

The rail freight data were provided by the Association of American Railroads (AAR) (2021) and were reported by individual U.S. Class I railroads and consolidated by the AAR. For clarity, "*freight volume*" refers to the weekly average IM units (for IM rail freight) or weekly average carloads (for commodity rail freight) of U.S. Class I railroads, which are railroads with annual revenues of at least $900 million in 2020 (Surface Transportation Board, 2022) excluding the U.S. operations of Canadian Pacific, Canadian National, and Groupo México Transportes. The archive of publicly available versions of the dataset and detailed methodologies and commodity classification can be accessed online (Association of American Railroads, 2021). Detailed description and semi-quantitative analysis of freight volume changes during the two disruptions, as well as their correlations with various economic metrics can be found in (Schofer et al., 2021).

Economic indicators as covariates were selected from a pool of relevant metrics based on the following criteria. As rail infrastructure is mostly subject to railroad investment and labor deployment may be slow to change because of labor contracts and challenges in hiring, the rail freight supply is relatively constant and inelastic in short term. Therefore, economic metrics should contribute to the models by reflecting demands for transportation services and show strong correlation with the freight volume changes. However, they should not be directly dependent on rail freight (*e.g.* coal production estimate and coal rail freight volume) for the lack of insights into the relationship between rail freight and the general economy. They should be publicly available and measured consistently and accurately, so that the established association with freight can be assessed and utilized in the future.

The economic indicator associated with IM was Real Personal Consumption Expenditures - Durable Goods (PCE) (U.S. Bureau of Economic Analysis, 2021)*,* and for coal as Industrial Production: Total Index (Not Seasonally Adjusted) (IP) (Board of Governors of the Federal Reserve System (US), 2021). Table 2 listed other variables assessed yet not chosen.

**Table 2 - Selected and Candidate Economic Indicators**

| Selected Economic Indicators | Candidate Economic Indicators |
|---|---|
| Real Personal Consumption Expenditures - Durable Goods (PCE) | Gross Domestic Product |
| Industrial Production: Total Index | Gross Output by Industry |
| | Individual components of Industrial Production (*e.g.* Manufacturing) |
| | Other Major Types of Products of Personal Consumption Expenditure |
| | Coal Production Estimate |
| | Field Production of Crude Oil |
| | Retail Sales: Retail and Food Services, Total, Nonstore Retailers, and Electronic Shopping and Mail-order Houses |
| | E-Commerce Retail Sales as a Percent of Total Sales |
| | Crude Oil Prices: West Texas Intermediate (WTI) - Cushing, Oklahoma |
| | Henry Hub Natural Gas Spot Price |
| | Unemployment Rate |

Nine types of IM and commodity shipment were covered, accounting for around 90% of rail freight volume in terms of carload and IM units. They are: IM; coal; chemical; grain; crushed stone, sand and gravel ("*sand*"); motor vehicles and equipment ("*auto*"); petroleum products ("*petroleum*"); metals and products ("*metal*"); and lumber and wood products ("*lumber*"). IM and coal, which covered around two-thirds of the total rail freight volume, were further explored under Scenarios 2 and 3 introduced in Section 1.

To maintain consistency of seasonal fluctuations across years and align with the aggregation of economic metrics, the weekly rail freight data were down-sampled to monthly by arithmetic averages. As discussed above, seasonal variations of data were identified as annual for those which had not been seasonally adjusted, *i.e.* the number of time steps is 12 for most cases.

**Figure 1** in Section 1 summarizes the three scenarios used in the ARIMA models to create a counterfactual baseline.

Scenario 1 is the trend continuation for IM and the eight commodities, projecting the freight volumes as if the disruption had not happened. It serves as a baseline to be compared with the actual freight volume for quantifying the impact of economic disruptions.

To illustrate the improvement in accuracy, freight-economy association, and usage of covariate-adapted models, Scenarios 2 and 3 were constructed for IM and coal freight. Scenario 2 was based on Scenario 1 with added economic indicators as covariates to the models so as to improve prediction performance, under the assumption that the freight and corresponding economic metrics move together. First, the main freight models were trained by the pre-disruption time series of freight and economic indicators. Then, the trend of the economic indicators was projected in a separate model trained by pre-disruption data, and its projection during the disruption was then used as covariates in the main rail freight models. The improvement in fit quality would suggest whether these economic metrics are useful in freight model specifications. In contrast, Scenario 3 made use of the actual data series of covariates during the disruption to inform the main freight model, reflecting the economic situation in reality. Discrepancies between these "as if" scenario projection values and actual freight volume suggest changes in the pattern of association between freight and economic conditions, possibly due to factors such as short-term supply constraints on the freight market.

The duration of the Great Recession was identified as December 2007 - June 2009 (National Bureau of Economic Research, 2021). The input time frame was selected to capture general trends between recessions. While the COVID-19 recession was identified to take place in February - April 2020, major impacts on freight were observed in April - May 2020.

## 4. RESULTS AND DISCUSSION

The results of IM and coal rail freight are first discussed to illustrate the application and insights derived from the scenarios, followed by an overall discussion of all IM and commodities.

**Figure 3** illustrates how the model results should be interpreted with respect to the corresponding RQ for the COVID-19 pandemic example. Projection/forecast results refer to the model prediction median unless otherwise specified, with the 95% confidence interval shown in the transparent cone. All results are tabulated in **Table 3**Error! Reference source not found.. The MAPE and MAD for 1-step and 12-step prediction of the past data were calculated for the actual data in 2005-2008 and 2014-2019.

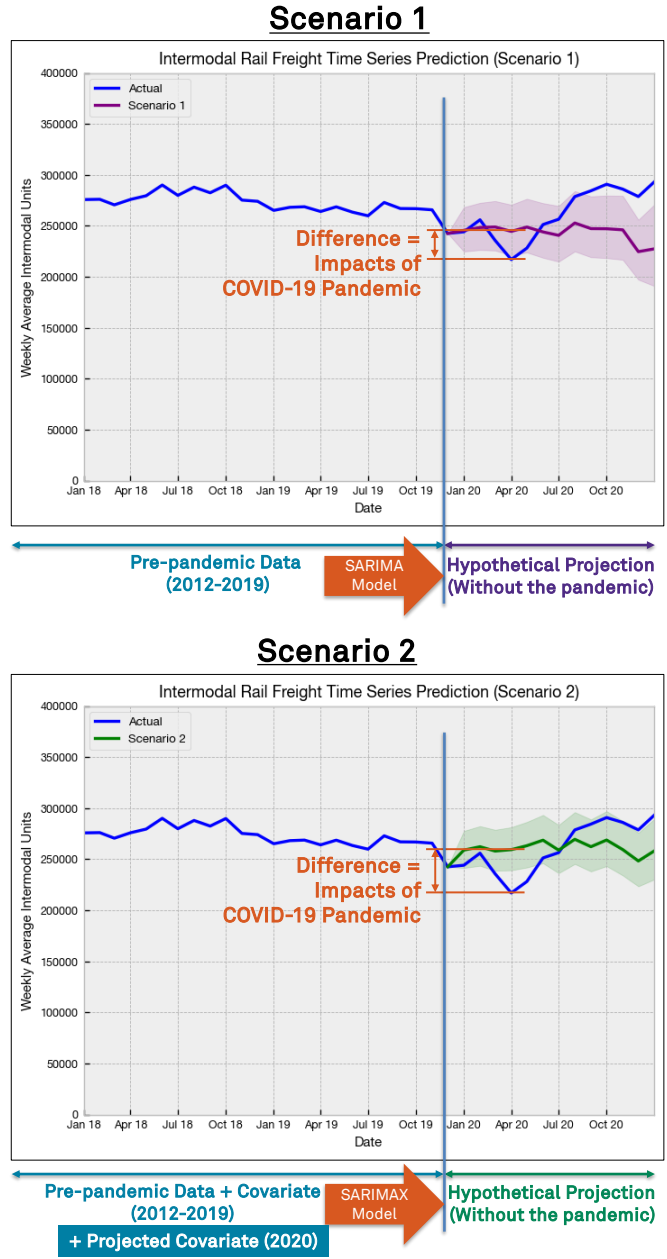

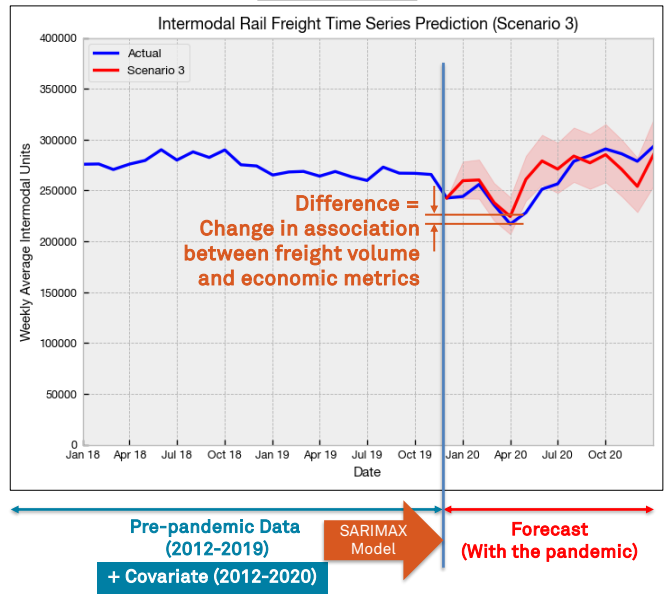

**Figure 3 – Illustration of Result Interpretation for the Three Scenarios**

**Table 3 - Parameters and Results of Model Training**

| Training Set Time Period | | 2012-2019 | | | | | | | | | | | | |
|---|---|---|---|---|---|---|---|---|---|---|---|---|---|---|
| Freight Component | | IM | | | Coal | | | Chemical | Grain | Sand | Auto | Petroleum | Metal | Lumber |
| Scenario | | 1 | PCE | 2&3 | 1 | IP | 2&3 | 1 | 1 | 1 | 1 | 1 | 1 | 1 |
| **Parameters** | P | 0 | 1 | 0 | 0 | 1 | 1 | 2 | 1 | 1 | 0 | 1 | 0 | 0 |
| | D | 1 | 1 | 1 | 1 | 1 | 1 | 1 | 1 | 1 | 1 | 1 | 1 | 1 |
| | Q | 1 | 0 | 1 | 0 | 0 | 0 | 0 | 2 | 1 | 1 | 0 | 0 | 1 |
| | P | 0 | / | 2 | 3 | 3 | 4 | 3 | 0 | 0 | 0 | 1 | 1 | 0 |
| | D | 1 | / | 1 | 1 | 1 | 1 | 1 | 1 | 1 | 1 | 1 | 1 | 1 |
| | Q | 0 | / | 0 | 0 | 1 | 0 | 1 | 1 | 1 | 1 | 0 | 0 | 1 |
| | S | 12 | / | 12 | 12 | 12 | 12 | 12 | 12 | 12 | 6 | 12 | 12 | 12 |
| | Transformation | log | / | log | / | / | log | log | log | log | sqrt | log | log | log |
| **Fitness** | AIC | -274 | 305 | -300 | -232 | 168 | -230 | -393 | -194 | -238 | 520 | -269 | -265 | -282 |
| | P>\|z\| for exog | / | / | 0.11 | / | / | 0.14 | / | / | / | / | / | / | / |
| | P(Q) | 0.88 | 0.83 | 0.83 | 0.8 | 0.9 | 0.96 | 0.81 | 0.86 | 0.97 | 0.37 | 0.9 | 0.92 | 0.93 |
| | 1-step MAPE | 3.29% | 0.65% | 2.75% | 4.33% | 0.50% | 4.34% | 1.55% | 5.44% | 4.07% | 5.17% | 3.74% | 3.67% | 3.15% |
| | 1-step MAD | 8506 | 0.86 | 7165 | 3766 | 0.50 | 3765 | 480 | 1175 | 906 | 849 | 459 | 337 | 105 |
| | 12-step MAPE | 5.75% | 1.27% | 4.20% | 11.43% | 1.99% | 9.83% | 2.31% | 11.39% | 9.12% | 5.97% | 23.19% | 11.88% | 7.74% |
| | 12-step MAD | 15014 | 1.69 | 10972 | 9687 | 2.00 | 8324 | 719 | 2432 | 2025 | 984 | 2928 | 1068 | 257 |
| **Estimated coefficients** | Intercept | / | 0.86 | / | / | / | / | / | / | / | / | / | / | / |
| | Exog | / | / | 0.00 | / | / | 0.01 | / | / | / | / | / | / | / |
| | AR1 | / | -0.34 | / | / | -0.22 | -0.08 | -0.43 | 0.55 | 0.51 | / | 0.22 | / | / |
| | AR2 | / | / | / | / | / | / | -0.39 | / | / | / | / | / | / |
| | MA1 | -0.67 | / | -0.66 | / | / | / | / | -0.58 | -0.78 | -0.79 | / | / | -0.45 |
| | MA2 | / | / | / | / | / | / | / | -0.30 | / | / | / | / | / |
| | ARS1 | / | / | -0.70 | -0.64 | -0.33 | -0.83 | -0.29 | / | / | -0.85 | -0.50 | -0.38 | -0.98 |
| | ARS2 | / | / | -0.34 | -0.53 | -0.39 | -0.70 | -0.29 | / | / | / | / | / | / |
| | ARS3 | / | / | / | -0.47 | -0.34 | -0.70 | -0.40 | / | / | / | / | / | / |
| | ARS4 | / | / | / | / | / | -0.36 | / | / | / | / | / | / | / |
| | MAS1 | / | / | / | / | -0.68 | / | -0.76 | -0.77 | -0.88 | / | / | / | / |

| Training Set Time Period | | 2002-2007 | | | | | | | | |
|---|---|---|---|---|---|---|---|---|---|---|
| Freight Component | | IM | Coal | Chemical | Grain | Sand | Auto | Petroleum | Metal | Lumber |
| Scenario | | 1 | 1 | 1 | 1 | 1 | 1 | 1 | 1 | 1 |
| Parameters | P | 2 | 1 | 1 | 0 | 2 | 1 | 2 | 1 | 1 |
| | D | 1 | 1 | 1 | 1 | 1 | 1 | 1 | 1 | 1 |
| | Q | 1 | 1 | 1 | 0 | 1 | 1 | 0 | 0 | 0 |
| | P | 0 | 1 | 0 | 1 | 0 | 1 | 1 | 1 | 1 |
| | D | 0 | 0 | 1 | 0 | 1 | 1 | 1 | 1 | 1 |
| | Q | 1 | 0 | 1 | 0 | 1 | 0 | 0 | 0 | 0 |
| | S | 12 | 12 | 12 | 12 | 12 | 12 | 12 | 12 | 12 |
| | Transformation | log | log | log | log | log | sqrt | log | log | log |
| Fitness | AIC | -245 | -299 | -270 | -202 | -161 | 334 | -230 | -206 | -199 |
| | P>\|z\| for exog | / | / | / | / | / | / | / | / | / |
| | P(Q) | 0.94 | 0.81 | 0.91 | 0.84 | 0.98 | 0.8 | 0.86 | 0.84 | 0.78 |
| | 1-step MAPE | 2.72% | 2.16% | 1.41% | 4.40% | 4.18% | 4.42% | 2.10% | 2.84% | 3.36% |
| | 1-step MAD | 6258 | 2993 | 415 | 987 | 883 | 898 | 128 | 378 | 165 |
| | 12-step MAPE | 4.33% | 2.54% | 4.06% | 6.62% | 8.03% | 5.15% | 3.07% | 7.02% | 13.11% |
| | 12-step MAD | 10063 | 3535 | 1197 | 1471 | 1687 | 1037 | 189 | 920 | 624 |
| Estimated coefficients | Intercept | / | / | / | / | / | / | / | / | / |
| | exog | / | / | / | / | / | / | / | / | / |
| | AR1 | -1.09 | 0.42 | 0.48 | / | -0.98 | 0.46 | -0.35 | -0.10 | -0.13 |
| | AR2 | -0.51 | / | / | / | -0.49 | / | -0.36 | / | / |
| | MA1 | 0.62 | -0.90 | -0.78 | / | 0.60 | -0.97 | / | / | / |
| | ARS1 | / | 0.98 | / | 0.65 | / | -0.27 | -0.66 | -0.71 | -0.22 |
| | MAS1 | 0.60 | -0.83 | -0.63 | / | -0.55 | / | / | / | / |

## 4.1 Intermodal (IM) Rail Freight

### 4.1.1 Scenario 1 - Trend Continuation

A model of *SARIMA(0,1,1,0,1,0,12)* with logarithmic transformation was constructed for the COVID-19 disruption with IM freight volume in 2012-2019 as input. The 1-month and 12-month MAPEs are 3.29% and 5.75% respectively, demonstrating a reasonable fitting quality. In **Figure 4,** the projected trend of Scenario 1 in 2020 extends the general decreasing trend in 2018-2019, while keeping the seasonal pattern, such as year-end drop. In April 2020, the actual IM freight was around 10% lower than the hypothetical baseline, highlighting the impact of initial lockdown measures for the coronavirus. In the second half of the year, actual freight rose to more than 15% higher than Scenario 1, as fueled by the shift in consumers' purchase of goods from experiences and the growth of e-commerce.

It is worth noting that this measured impact (nearly 30% at the end of 2020) was greater than simple comparison with pre-disruption level or year-on-year comparison, due to the pre-COVID decreasing trend in freight volume.

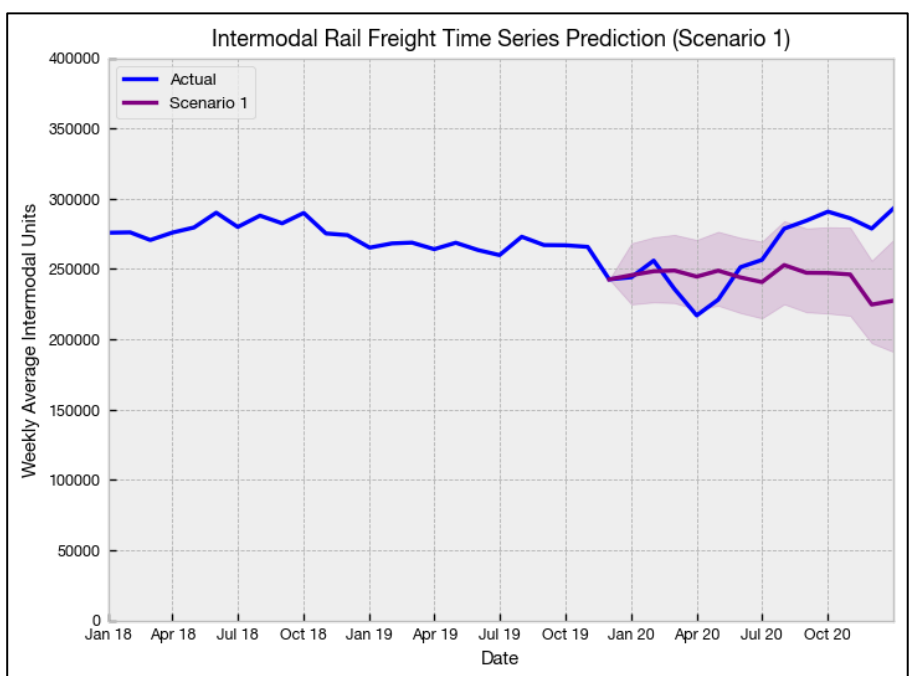

**Figure 4 – IM Rail Freight Projection (Scenario 1) (2018-2020)**

For the IM freight during the Great Recession, a *SARIMA(2,1,1,0,0,1,12)* model with logarithmic transformation was adopted. The 1-month and 12-month MAPEs are 2.72% and 4.33% respectively, also suggesting good fitting. **Figure 5** shows the result during the Great Recession with a significantly different pattern. The IM freight fluctuated around the projected volume until falling to around 20% under the projection in December 2008. This illustrates the difference in the nature between the supply-side shocks in 2020 and the demand-side recession in 2008. The former is relatively abrupt changes due to shortage of goods, while the latter shows more gradual reduction in freight volume following the drop in consumers' demand, retailers' order, and thereby freight demand.

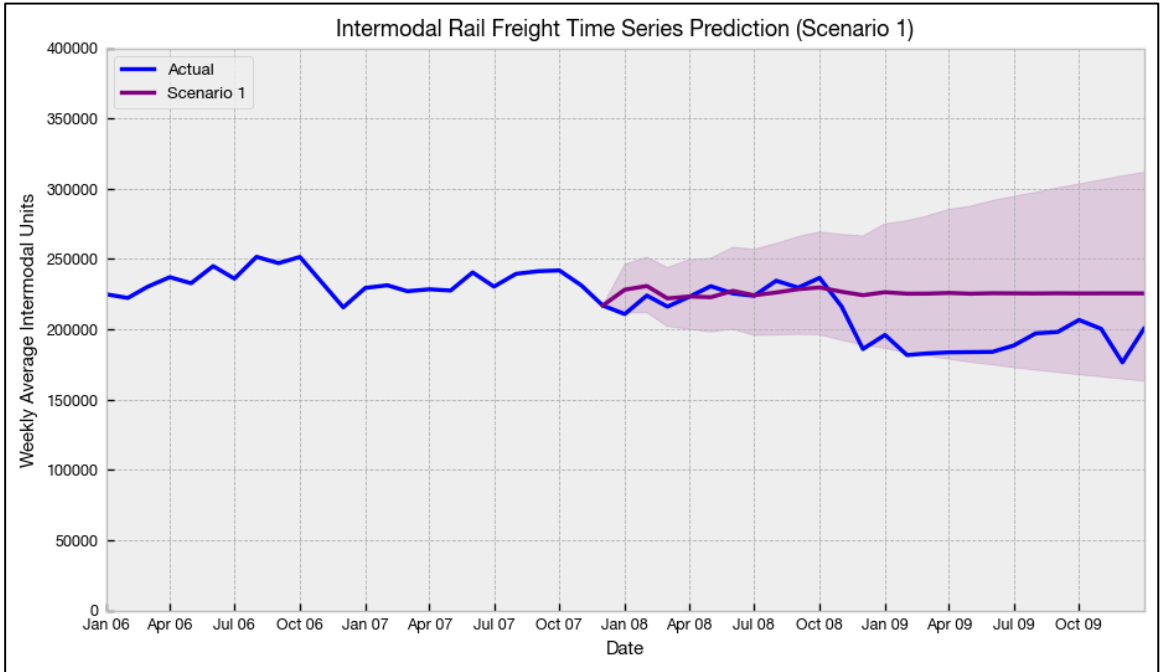

**Figure 5 – IM Rail Freight Projection (Scenario 1) (2006-2009)**

*4.1.2 Covariate for Scenarios 2 and 3*

The economic indicator – PCE showed a close correlation to IM freight. **Figure 6** shows both time series as percentage change from 2008, indicating general growing trends. During the financial crisis in 2008, PCE dropped by 13% from August to October 2008 while IM freight decreased by 26% from October 2008 to January 2009. Both started to recover in March 2009. Similarly, during the COVID-19 lockdown in 2020, PCE and IM freight fell by 22% and 17% respectively from February to April 2020 due to the short-term supply disruption. However, both fully recovered within three months and grew 10% higher than pre-pandemic level in the third quarter of 2020, showing rapid rebounds.

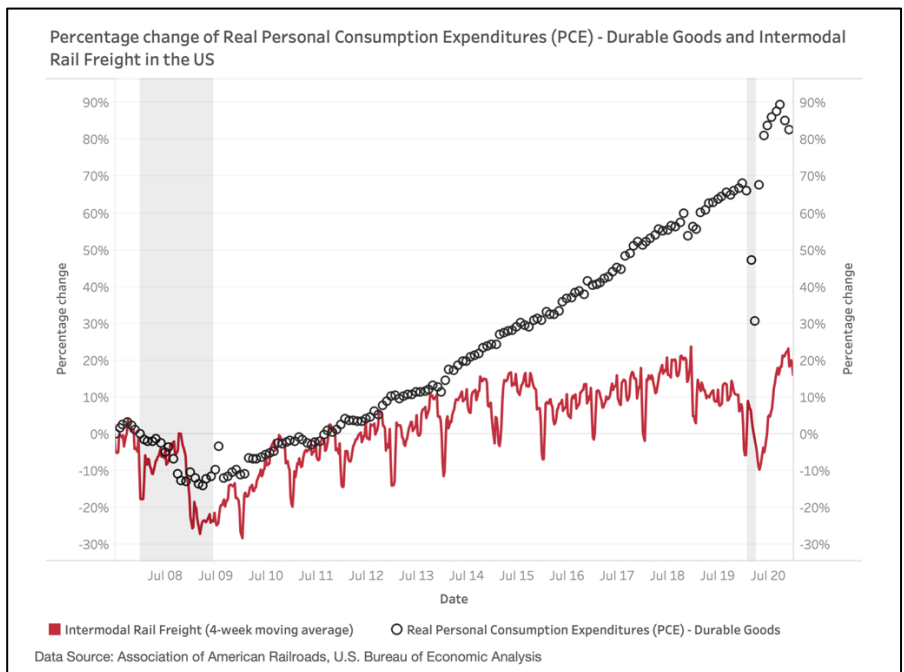

**Figure 6 – Percentage change of Real Personal Consumption Expenditure (PCE) – Durable Goods and IM Rail Freight in the US (2007-2020)**

**Figure 7** shows the scattered plot for 2012-2019 in blue dots and 2020 in green dots. The correlation coefficients are 0.66 before the pandemic and 0.85 during the pandemic. The strong correlation corroborates the use of PCE as a covariate for the IM freight model.

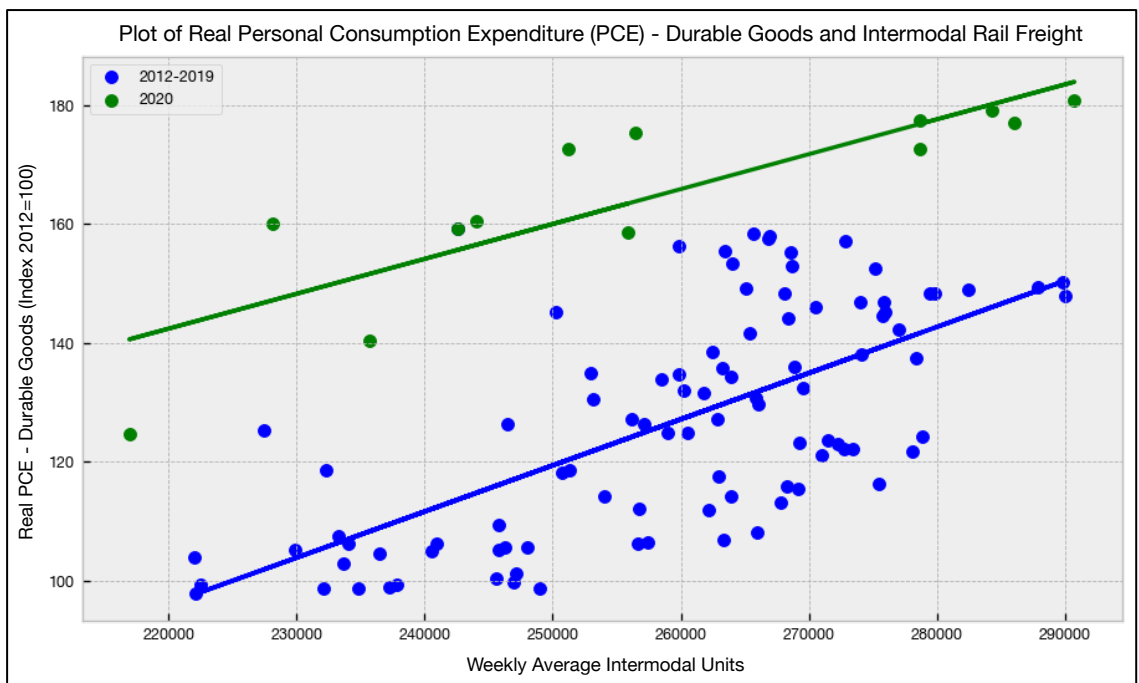

**Figure 7 – Covariate and IM Rail Freight (2012-2020)**

The time series model for the PCE was *ARIMA(1,1,0)* with 1-step MAPE 0.65% and 12-step MAPE 1.27%. Because the PCE data published by Federal Reserve had been seasonally adjusted, no seasonal components were needed. In **Figure 8**, the projection from the models extended past trends as straight lines because the measure does not include any seasonal fluctuation. While consumption was suppressed when the pandemic first struck in Apr 2020, after the initial lockdown ended, the PCE rose quickly and was nearly 10% higher than the projection, which illustrates a jump in personal consumption during the second half of 2020.

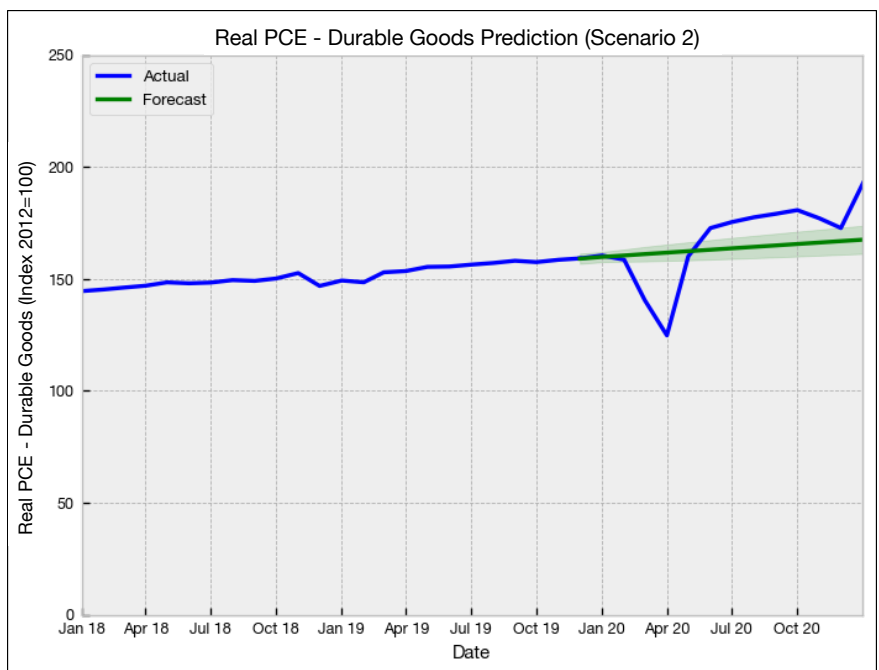

**Figure 8 – Projection for Real Personal Consumption Expenditure (PCE) – Durable Goods (2018-2020) (Scenario 2)**

*4.1.3 Scenario 2 – Covariate-adapted Trend Continuation*

By taking the projected PCE as a covariate, a time series model for Scenarios 2 and 3 was constructed as *SARIMAX(0,1,1,2,1,0,12)*. The 1-step and 12-step MAPE are 2.75% and 4.20% respectively, which are smaller than those in Scenario 1. The difference manifests the improvement in fitting by incorporating the covariate.

The projection of Scenario 2 in **Figure 9** shows a general decrease similar to Scenario 1, though less drastic. This can be explained by the projected increase in PCE used in the main time series model. The largest difference from the actual volume was as much as -15%. Subsequently, IM freight caught up in July and rose higher in the second half of the year.

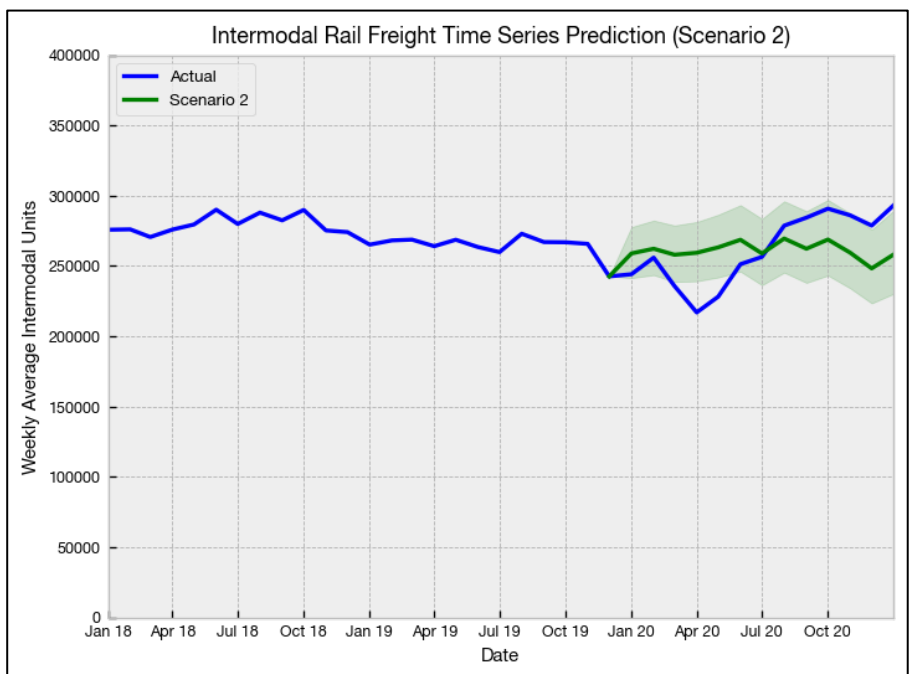

**Figure 9 – IM Rail Freight Projection (Scenario 2) (2018-2020)**

*4.1.4 Scenario 3 - Actual Covariate-adapted Forecast*

For Scenario 3 using actual covariate data during the pandemic, the forecasted freight shown in **Figure 10** dropped sharply in April 2020 as in reality. However, it bounced back quickly in May and June, which led the actual volume by more than one month. This difference demonstrates the change in association between the two variables, potentially due to the fact that demand recovery was led by an increase in consumption. The IM freight could not immediately catch up with this swift change from the initial lockdown stage, probably due to supply constraints such as labor shortages because of health concerns.

During the recovery stage, actual freight volumes persisted above the forecasted level, aligning with the observed shift in the consumption pattern (from buying experiences to products), which drove higher IM demand.

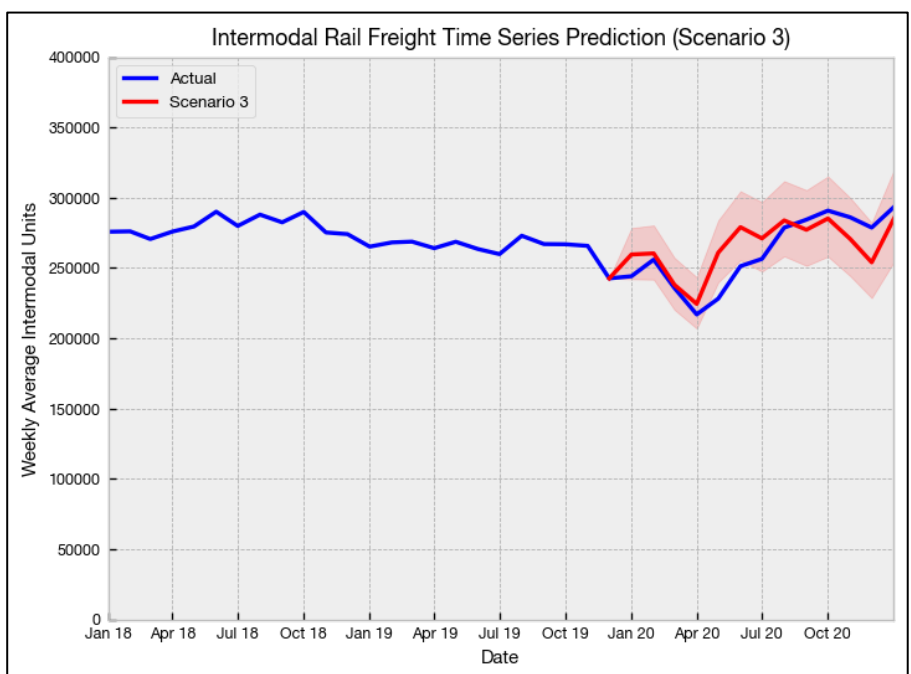

**Figure 10 – IM Rail Freight Forecast (Scenario 3) (2018-2020)**

## 4.2 Coal Rail Freight

*4.2.1 Scenario 1 - Trend Continuation*

    **Figure 11** shows the results of the model for coal freight, *SARIMA(0,1,0,3,1,0,12)* with MAPE 4.33% for 1-month and 11.43% for 12-month projection. There was a general decreasing trend in 2020, led by the prevailing decline of coal usage due to market-driven shifts of input energy to natural gas and renewable energy. While both the projected and actual volume decreased in January - March 2020, the actual traffic decreased further to more than 25% below the projected level. This captured the unforeseen lockdown restrictions, in particular those imposed on industries, and the energy price crash. Although coal freight recovered afterward, it remained lower than the trend continuation projection until the end of 2020. However, the measured decrease was smaller than simple comparison with pre-COVID level or year-on-year comparison, after considering the secular drop in coal freight volume.

    The model for the Great Recession was chosen as *SARIMA(1,1,1,1,0,0,12)* with MAPE 2.16% for 1-month and 2.54% for 12-month projection. In **Figure 12**, similar to IM, coal freight showed no sign of decrease and even exceeded the projection of trend continuation when the recession unfolded in 2008. However, in May 2009, the actual volume dropped to as low as 15% below the projection level. An even lower trough was observed in December 2009. The difference likely reflected the reduction in energy demand caused by the slowdown in economic activities, including industrial, production during the Great Recession.

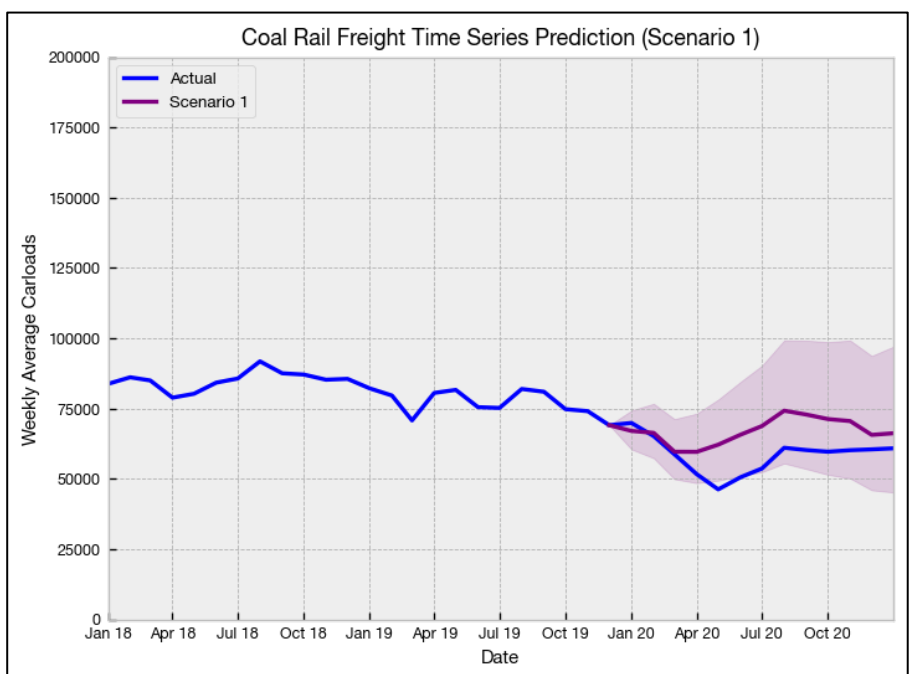

**Figure 11 – Coal Rail Freight Projection (Scenario 1) (2018-2020)**

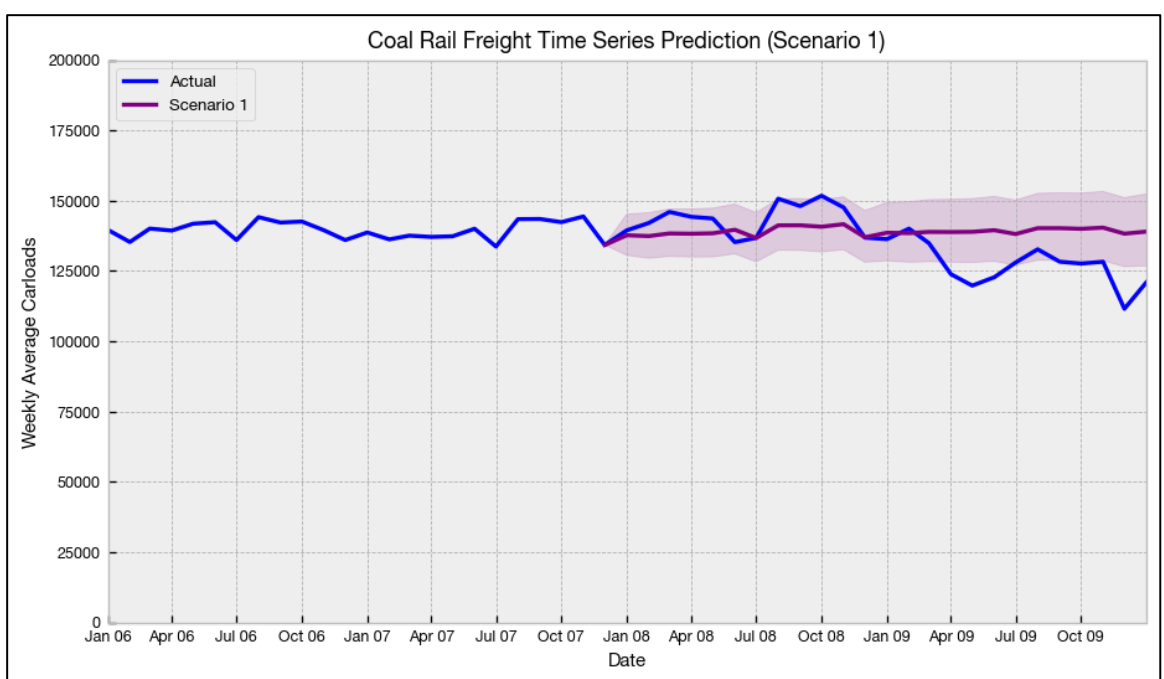

**Figure 12 – Coal Rail Freight Projection (Scenario 1) (2006-2009)**

### 4.2.2 Covariate for Scenarios 2 and 3

IP was chosen as the covariate of coal rail freight for the role of industrial activities in energy consumption. **Figure 13** shows the past pattern of both measures. A secular decline of 45% in coal rail freight from 2009 to 2020 is observed, which can be attributed to the competition from natural gas and increasing production of oil in the U.S. However, its short-term fluctuations still match with IP. During the recession in 2008, coal freight decreases by 10% from March to May 2009 following the 17% drop in IP in the first quarter. Similarly, during the early stage of COVID-19 pandemic, coal freight volume decreased by more than 30%, subsequent to the 20% fall of IP from February to April 2020. A V-shape rebound was observed in both time series, which recovered half of the losses by June 2020.

To account for the declining trend of coal freight in the period, first-order seasonal differencing was carried out. The correlation coefficient of seasonal difference of IP and rail freight is 0.69 in 2012-2020, indicating a strong correlation after first-order seasonal differencing (**Figure 14**).

**Figure 15** shows the model result for IP with *SARIMA(1,1,0,3,1,1,12)*. In April 2020, actual IP dropped by 15% below the hypothetical baseline of Scenario 1, depicting the impact of the initial lockdown, and did not recover fully in the year.

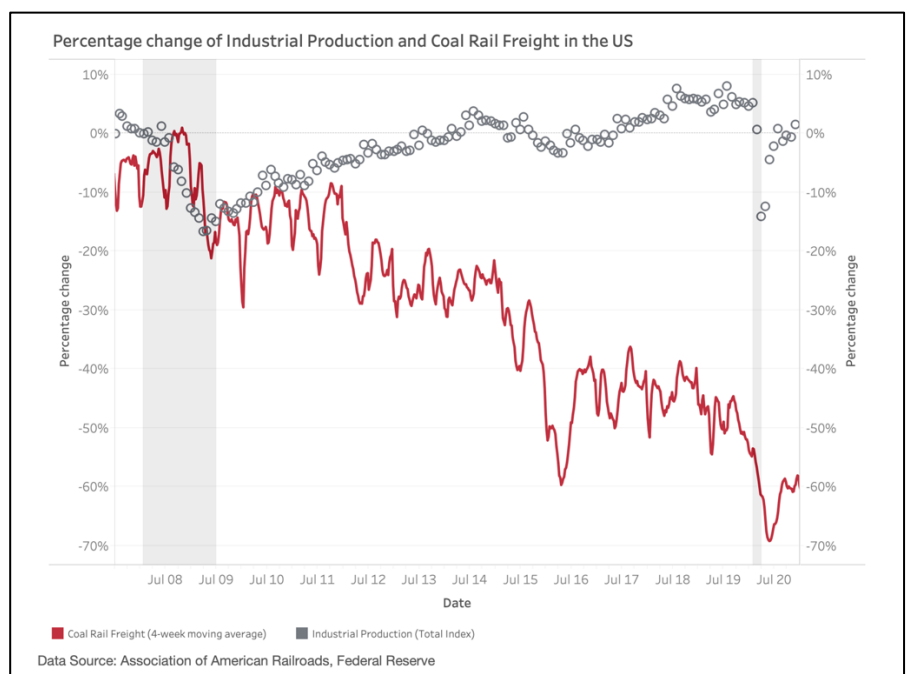

**Figure 13 - Percentage change of Industrial Production and Coal Rail Freight in the US (2008-2020)**

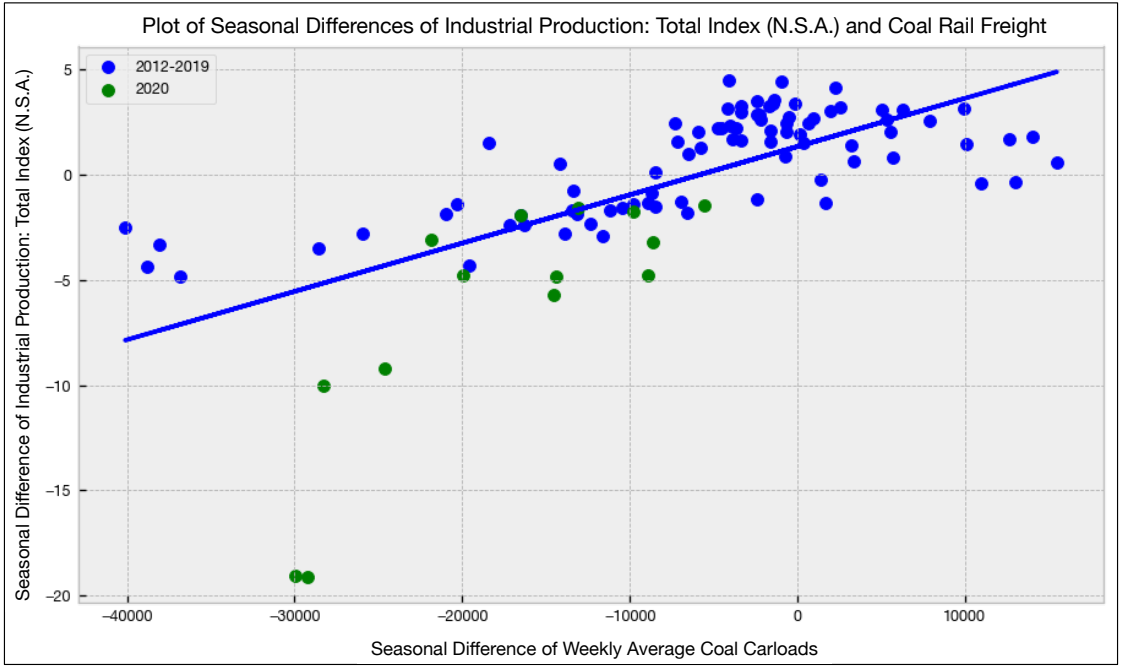

**Figure 14 - Seasonal Differences of Covariate and Coal Rail Freight (2012-2020)**

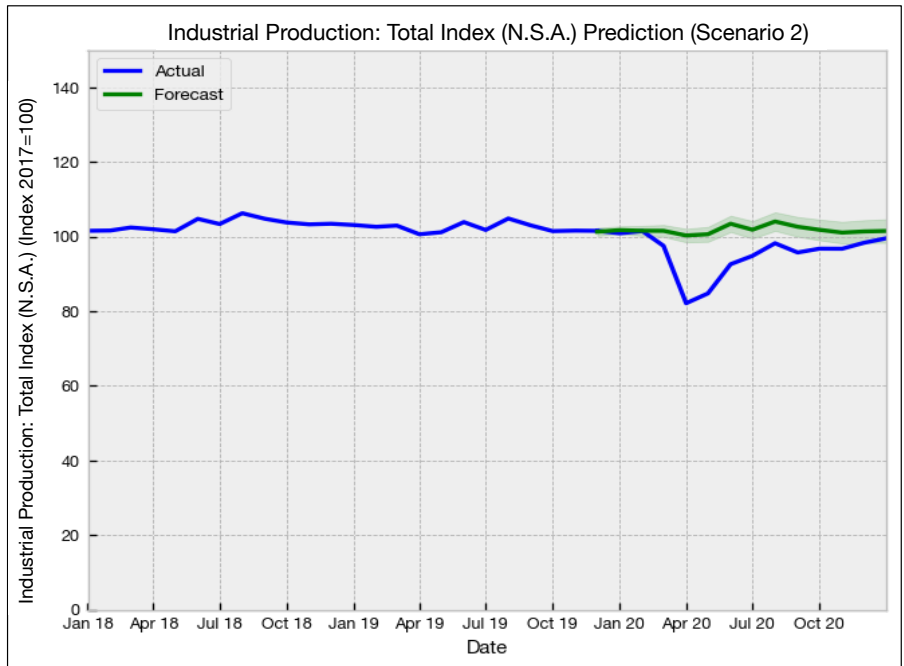

**Figure** 15 **– Projection for Industrial Production (IP): Total Index (2018-2020)**

*4.2.3 Scenario 2 - Covariate-adapted Trend Continuation*

   Incorporating the output from the IP estimate, **Figure 16** shows the projection of coal rail freight from the model *SARIMAX(1,1,0,4,1,0,12)*, which is slightly lower than Scenario 1. The fitting quality of the model was improved with the use of covariates, which cut the 12-month

MAPE from 11.43% to 9.83%. The resulting impact measured at the peak of the pandemic was found to be as high as -25%.

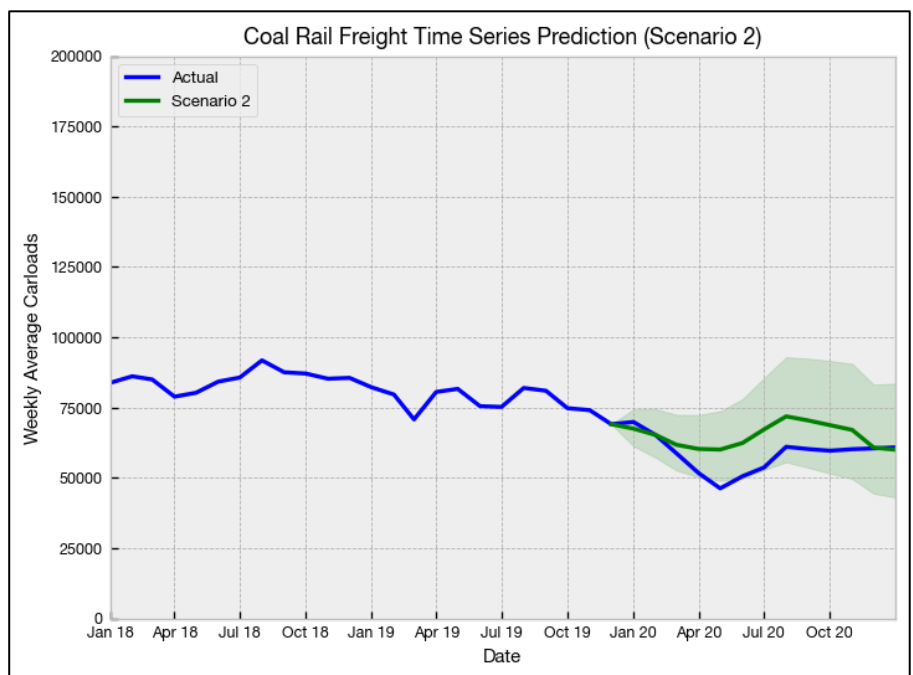

**Figure 16 – Coal Rail Freight Projection (Scenario 2) (2018-2020)**

*4.2.4 Scenario 3 - Actual Covariate-adapted Forecast*

Taking into account the actual IP drop in April-May 2021, the forecasted result of Scenario 3 in **Figure 17** was close to the actual freight volume, with time lead by around one month. This suggests that the interaction between coal rail freight and IP remained largely unchanged during the pandemic.

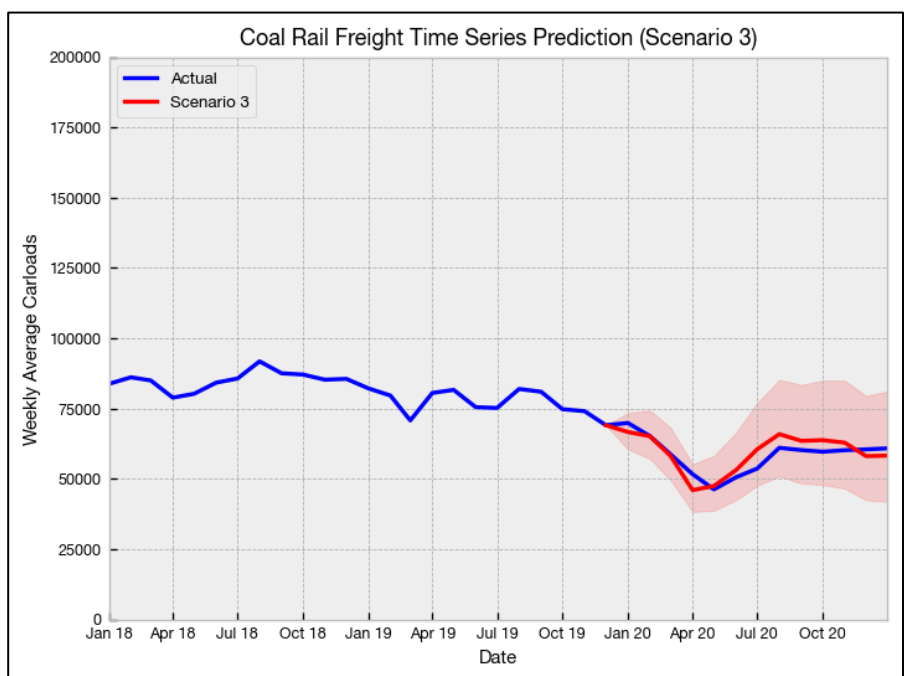

**Figure 17 – Coal Rail Freight Forecast (Scenario 3) (2018-2020)**

## 4.3 Overall Commodities (Scenario 1 - Trend Continuation)

The framework applied to IM and coal can be expanded to other commodities. To demonstrate the application, SARIMA models were used on the other seven types of commodities to construct Scenario 1. Instead of detailed discussion of individual results, their characteristics are summarized and compared in Recovery Pace Plots.

### 4.3.1 Recovery Pace Plots

**Figure 18** shows the Recovery Pace Plot for the COVID-19 pandemic in 2020, where x-axis and y-axis represent the magnitudes of disruption (Apr-May 2020) and recovery (Oct-Dec 2020) respectively in terms of ratios of actual freight volume to the counterfactual baseline in Scenario 1. Four regions were delineated by the x-axis, y-axis, and diagonal $y=x$. Region A is inapplicable in a recession. Region B indicates more freight activities than it would be without the disruption during the recovery phase, suggesting a strong rebound. Region C shows gradual recovery. Region D manifests further drop of freight volume during the recovery stage. In other words, Regions B, C, and D are in order of preferred outcome for recovery.

The classification of regions assists in identifying different recovery patterns of each commodity. In Region B, IM had the largest growth after the initial drop as previously discussed. Grain was only slightly impacted at the peak of the pandemic, but enjoyed a growth only slightly less than IM in the recovery phase. Similarly, lumber and chemical suffered 10%-25% drop in the second quarter due to production limitation, but they fully recovered by the year end after the major lockdown was lifted. At the lower end, Region C lies metal, which shared a similar pattern but a less significant peak drop, and auto, as an outlier in the chart, the drop of which can be explained by the factory shutdown during the initial wave of virus spread, causing an 80% decrease. Auto/auto parts freight volumes nearly fully recovered as the restrictions were lifted. Along the border of Regions C and D are commodities related to energy, namely coal, sand, and petroleum, which were greatly impacted by the reduction in energy usage since the pandemic. In particular, freight volumes of sand and petroleum in the recovery phase were even lower than those during the disruption.

A general trend was observed that components with a greater drop in the recession phase recovered less in the recovery stage in 2020, except auto. A best-fit line is drawn in blue by excluding auto. IM, lumber, and metal on the left of the line show faster recovery pace than other commodities.

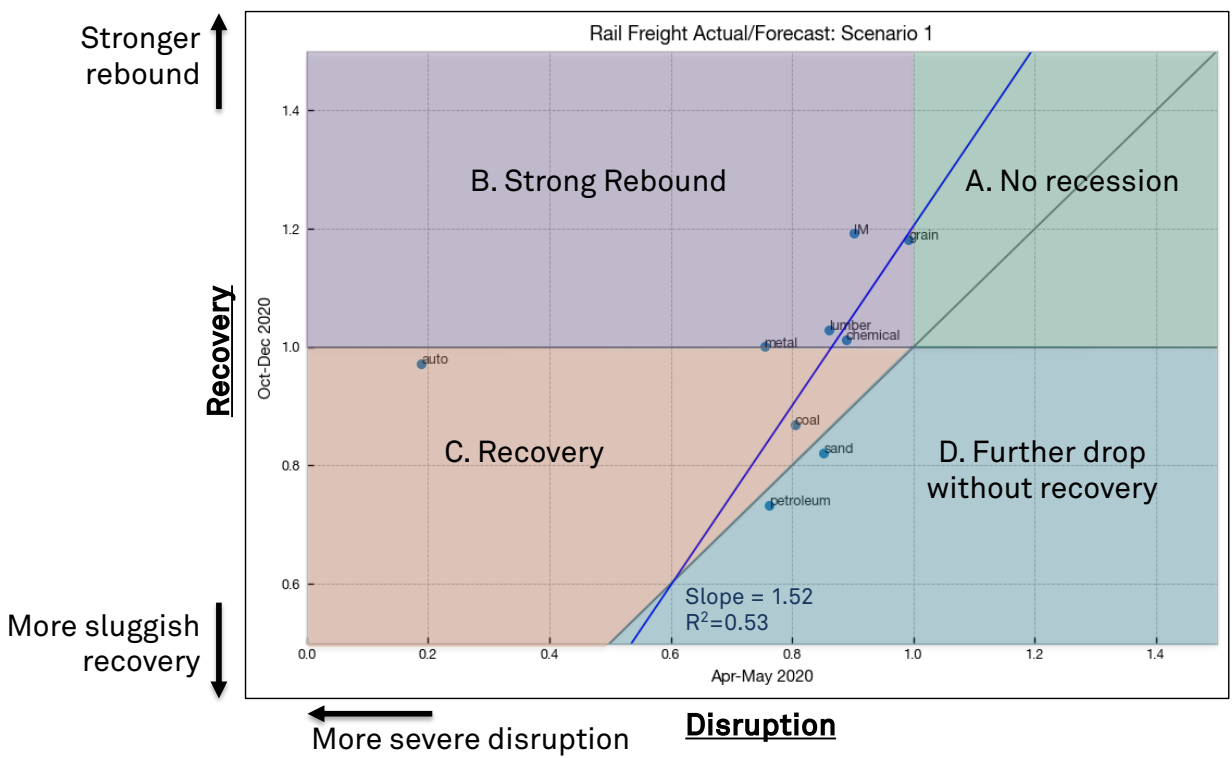

**Figure 18 – Recovery Pace Plot (2020)**

**Figure 19** shows the Recovery Pace Plot in 2008 for the Great Recession, in which the x-axis and y-axis represent the magnitudes of disruption (Q2) and recovery (Q4) respectively.

While the general trend of "greater drop, smaller recovery" persisted, the performance of IM and different commodities was more concentrated and in Region C. This, once again, highlights the difference in nature of the two disruptions. The Great Recession recorded a significant and long-lasting drop in demand, in particular for products, prompted by both demand and supply decline, leading to a slower recovery. The top group, grain and IM, showed moderate drop and faster recovery. Industrial products, such as lumber and chemical, were among the middle in drop and recovery. Energy-related commodities, such as coal, petroleum, and sand (along the border in Region D), were hard hit during the recession and recovered the slowest, recalling another crash in energy prices during the recession. Lastly, the auto freight volume was drastically smaller than the projection in the second quarter of 2009 due to the production and financial issues of motor vehicle companies following the financial crisis. Metals shared a similar trend.

The best-fit line in blue excluding rail freight of auto and metal has a higher $R^2$, showing the recovery paces of commodities are more concentrated compared to 2020. However, its smaller slope indicates an overall slower pace. Grain, IM, and lumber lying on the left hand side are the best performing commodities in recovery.

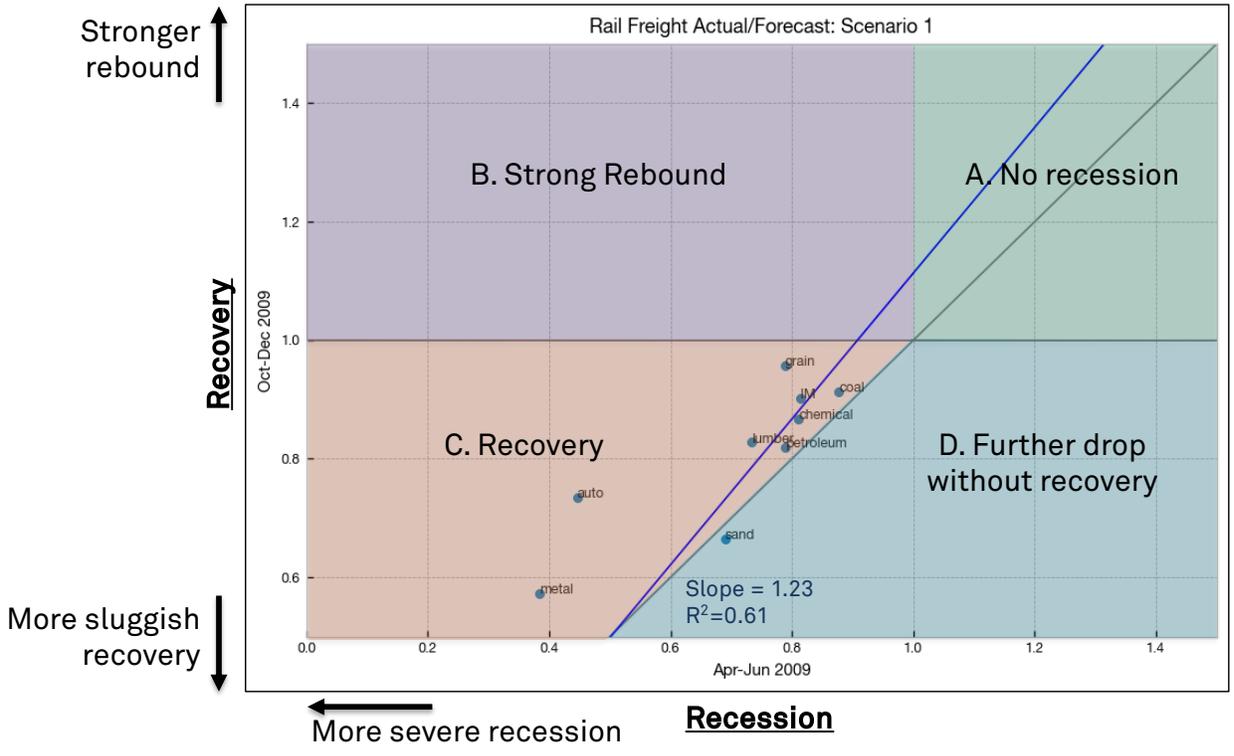

**Figure 19 –Recovery Pace Plot (2009)**

*4.3.2 Model Results*

The time series model results under the two economic disruptions for individual commodities are tabulated in **Table 4** under different categories of drop and recovery patterns.

**Table 4 - Rail Freight Projection (Scenario 1)**

| Year | 2018-2020 | 2006-2009 |
|------|-----------|-----------|
| **I. Moderate drop, faster recovery** | | |
| **IM** |  |  |
| **Grain** |  |  |

| Year | 2018-2020 | 2006-2009 |
|---|---|---|
| **II. Moderate drop, moderate recovery** | | |
| **Chemical** | 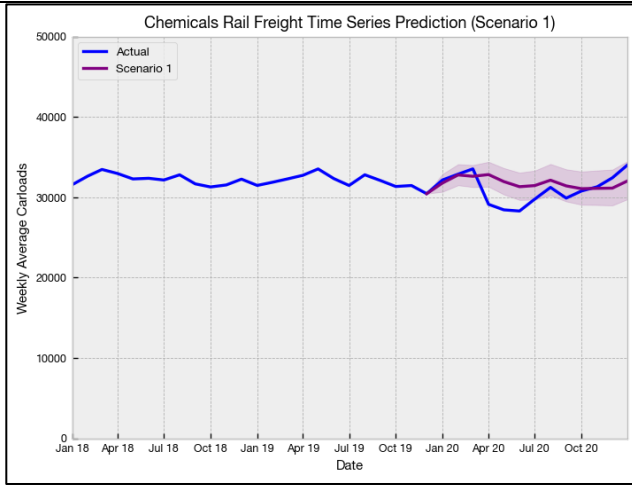 | 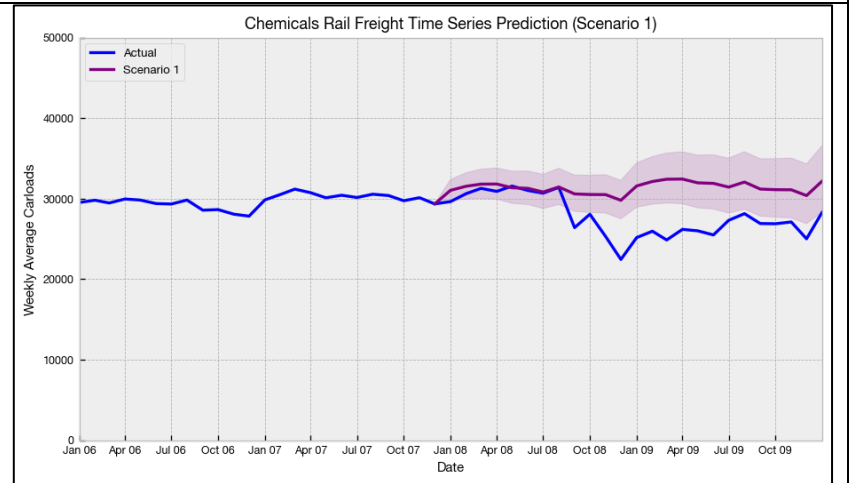 |
| **Lumber** | 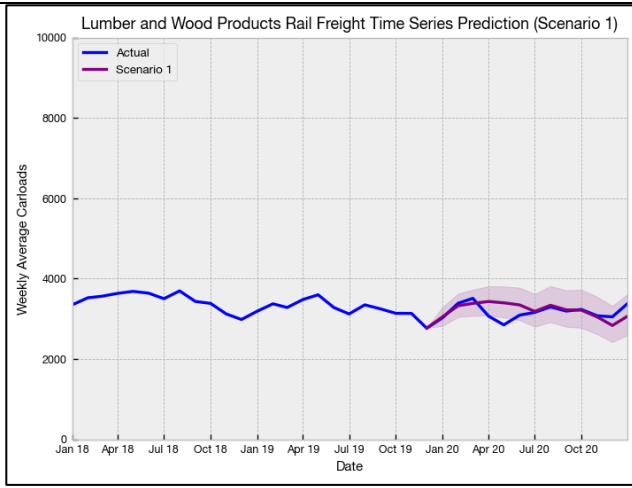 | 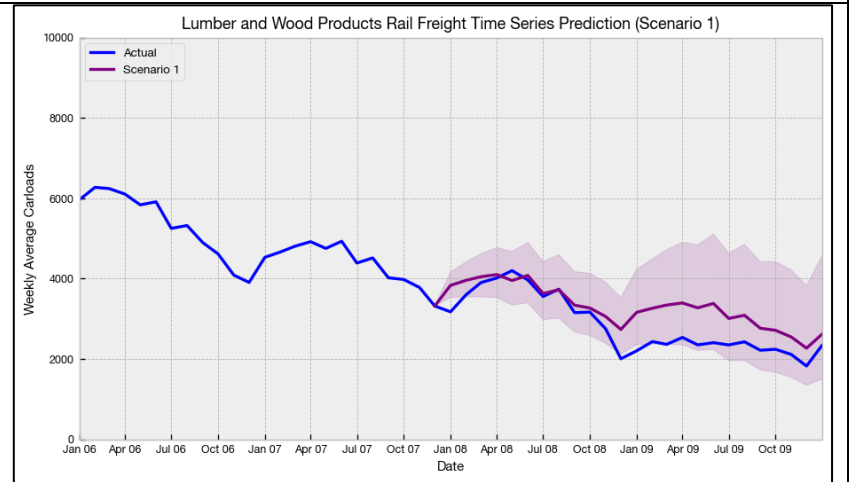 |

| Year | 2018-2020 | 2006-2009 |
|---|---|---|
| **III. Greater drop, sluggish recovery** | | |
| **Coal** | 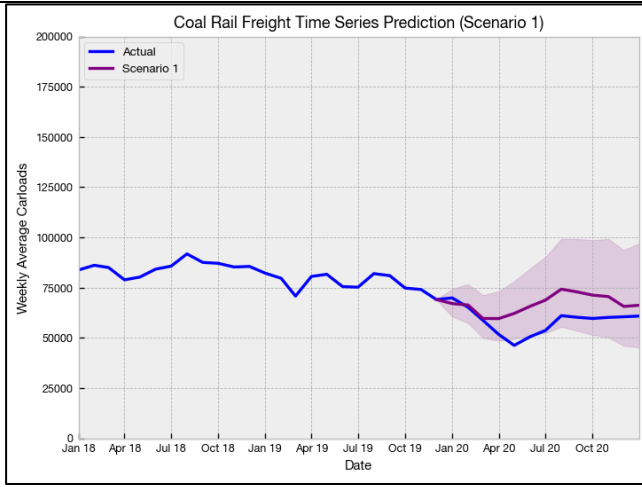 | 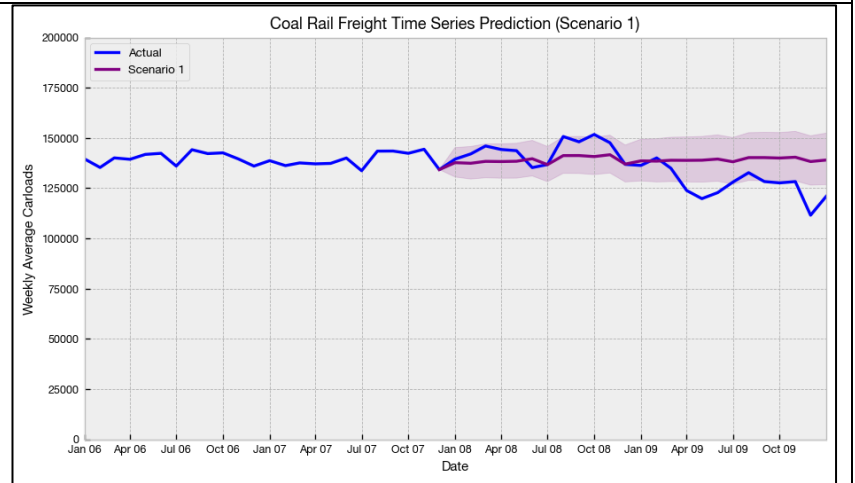 |
| **Oil** | 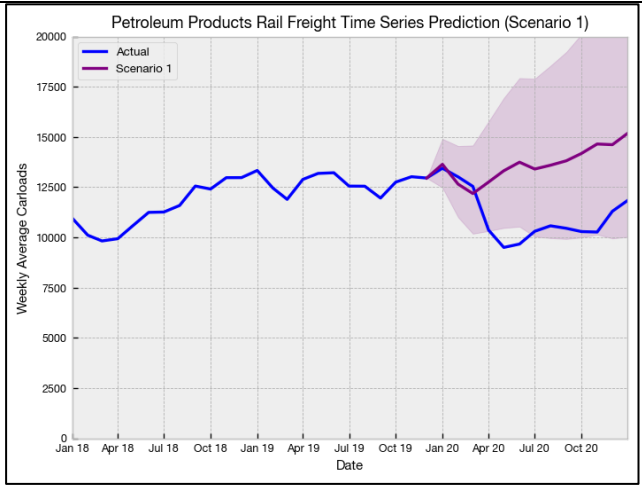 | 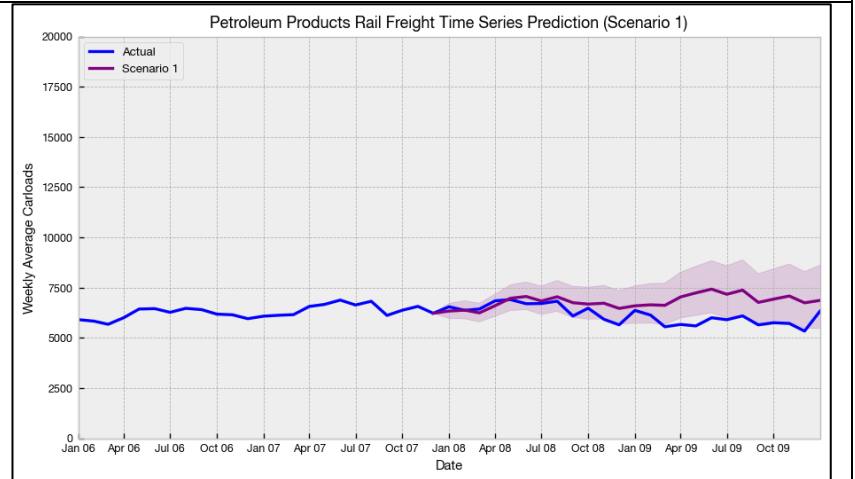 |

| Year | 2018-2020 | 2006-2009 |
|------|-----------|-----------|
| **III. Greater drop, sluggish recovery** | | |
| **Sand** | 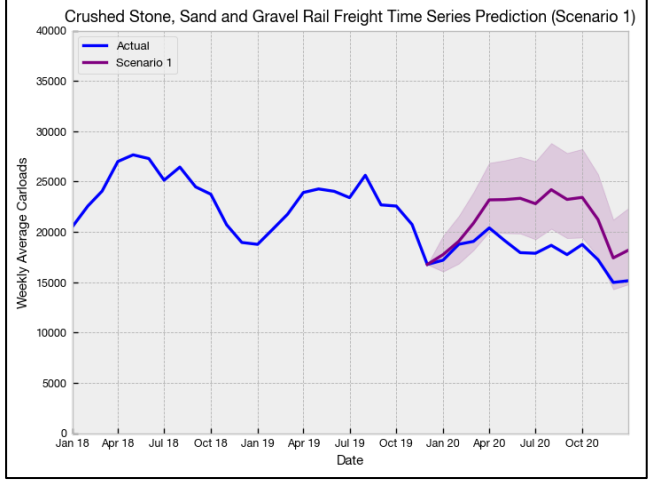 | 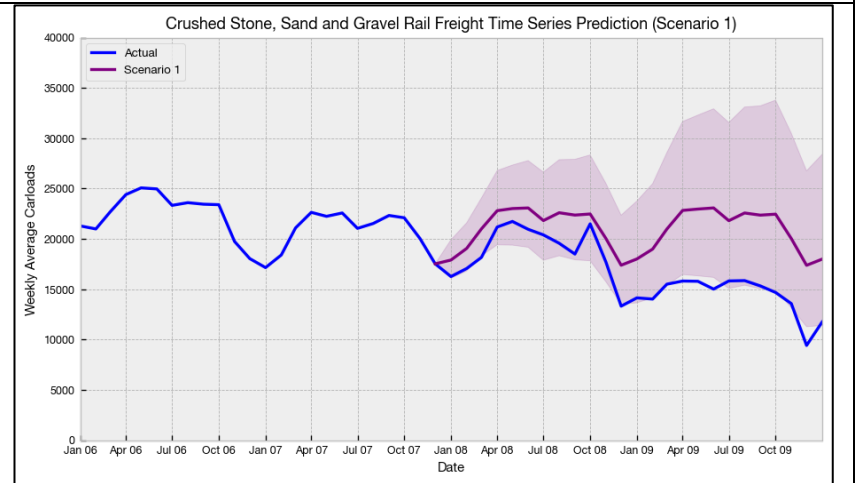 |

| Year | 2018-2020 | 2006-2009 |
|---|---|---|
| **IV. Sharpest drop, moderate recovery** | | |
| **Auto** | 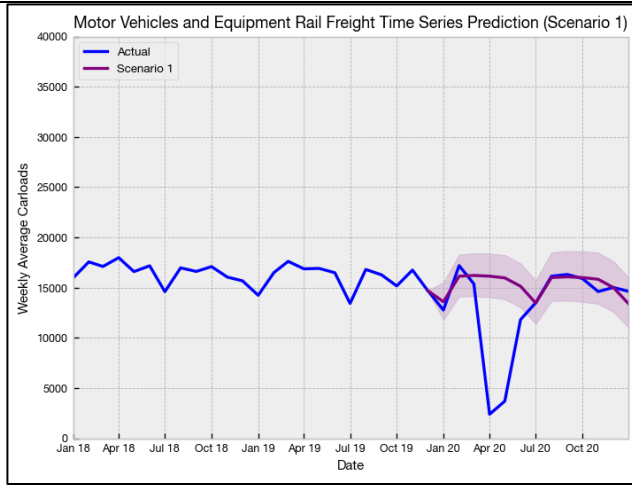 | 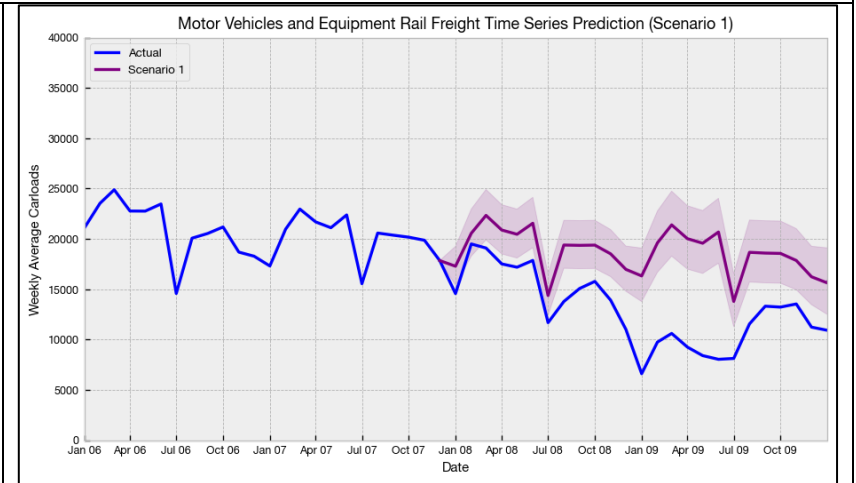 |
| **Metals** | 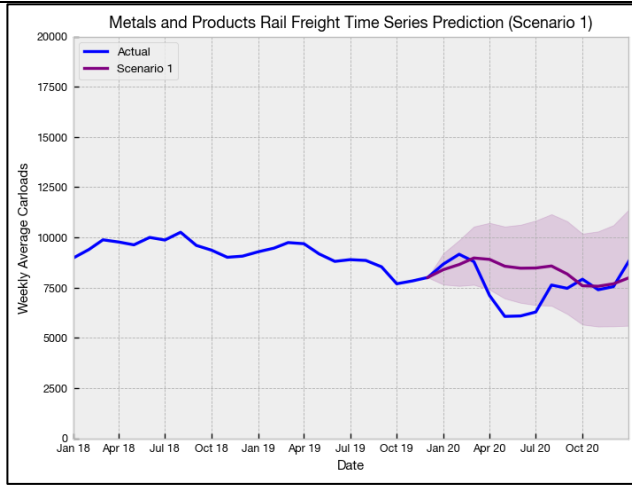 | 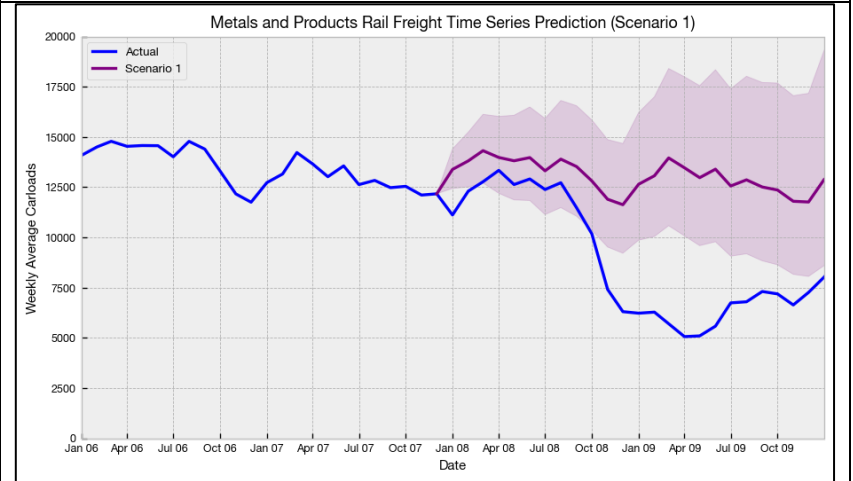 |

## 5. CONCLUSIONS

Rail is a core component of the logistics system in the U.S.. Rail freight volumes are affected by general trends, seasonal changes, random fluctuations, and the focus of this paper, major, short-term disruptions during economic recessions. The ARIMA family of models are formulated specifically for the problem addressed here - extending past trends. A framework of selecting parameters and evaluating model performance was utilized to build 22 SARIMA/SARIMAX models which projected/forecast three scenarios to provide counterfactual baselines against which to assess the impacts of economic disruptions in IM and eight commodities. The methodological and substantive findings are summarized in **Table 5**.

**Table 5 - Summary of Findings corresponding to the Research Questions (RQs)**

| RQ | Methodological Findings | Substantive Findings |
|---|---|---|
| 1. How can the disruption impacts be measured in rail freight volumes while separating freight activities from seasonality and structural trends? | • Scenario 1 (Trend continuation) was constructed with SARIMA models to project hypothetical baseline which captures seasonality and structural trends.<br>• Disruption impacts measured were different from those by simple comparison with pre-disruption levels or year-on-year depending on secular trend.<br>• Recovery Pace Plots support rapid comparison in rebound/recovery speed across freight components. | • IM freight volume showed significant disruption effect (+30%) at the end of 2020.<br>• IM and non-energy commodities showed stronger yet uneven rebound after COVID while slowly recovering after the Great Recession. |
| 2. Does the inclusion of broad economic metrics (covariates) in model specifications improve the ability to capture temporal patterns of rail freight volumes and lead to more meaningful pre- and post- pandemic comparisons? | • Scenario 2 (Covariate-adapted Trend Continuation) was constructed with SARIMAX main models. The inputs included projected economic covariates from separate ARIMA models.<br>• Improvements in prediction performance were observed. | • The impacts on IM and coal freight during the pandemic peak were as severe as -15% and -25%. |
| 3. Did the association between freight volumes and the underlying economic metrics (co-variates) established pre-pandemic continue to hold during the disruption and subsequent recovery? | • Scenario 3 (Actual Covariate-adapted Forecast) was constructed with SARIMAX main models. The inputs included forecasted economic covariates from separate ARIMA models. | • The IM freight association with PCE was maintained during the initial phase of the pandemic, but IM lagged behind during recovery.<br>• The coal freight association with industrial production was largely constant. |

With the examples of the two recessions, this paper has established and demonstrated a systematic approach using time series analysis to compare quantitatively the impacts of economic disruptions on rail freight. Scenario 1 modeled seasonal variations with continued underlying trends and provided the key basis for comparison with actual volumes to measure disruption impacts. The two disruption examples found the actual impacts to be more significant than the simple comparison with pre-disruption levels or year-on-year differences. The introduction of Recovery Pace Plots enables systematic evaluation and comparison of rebound/recovery patterns across IM and different commodities. Scenario 2 incorporated projected economic metrics as covariates. Results showed improvement in model prediction performance to observed pre-disruption data by capturing long term structural trends in addition to seasonality. Scenario 3 forecasted rail freight based on actual economic data during disruptions for comparison and analysis of impacting factors.

Substantive insights were drawn from the differences between the two disruptions. During the Great Recession, both supply and demand sides were heavily impacted, leading to a sluggish recovery in freight volumes. In contrast, the COVID-19 pandemic was a short-term supply shock followed by a positive demand shock, which led to a rapid albeit (+30% for IM) uneven recovery due to the ongoing pandemic and shift in purchase patterns. The differences in scale and pace of drop and recovery across freight components supported the prior hypothesis of uneven impacts, as seen from the Recovery Pace Plots. The IM example suggested possible lag of supply chain response to freight demand initially during the pandemic, while the association between coal and industrial production persisted.

Overall, the proposed approach can be used to enhance understanding of the disruptions themselves, interactions between supply and demand, and correlation with general economic trends, for example, the growth of e-commerce with IM traffic, which was gaining as a share of total rail freight volume. The approach is also useful to continue tracking the recovery pace, and to place it in context relative to previous existing trends. These can form a basis for setting expectations towards outcomes of large-scale disruptions to the (rail) freight system.

While rail and IM freight in the U.S. were used as examples, the proposed method can be applied to other modes, commodities, and scope. A potential direction to explore would be to identify the most promising economic indicators as covariates to capture trends and changes under different situations. Such insights will be beneficial to improving modeling and forecasting efforts of potential future disruptions, thereby informing the infrastructure planning, implementation, and operations strategy. This will help in enhancing readiness of the freight industry and boosting resilience in our supply chain.